# A mathematical discussion of Pons Viver's implementation of Löwdin's spin projection operator

Terutaka Yoshizawa

Department of Chemistry and Biochemistry, Graduate School of Humanities and Sciences, Ochanomizu University, 2-1-1 Otsuka, Bunkyo-ku, Tokyo 112-8610, Japan

Electronic mail: yoshizawaterutaka@gmail.com

**Abstract**

Recently the molecular electronic structure theories for efficiently treating static (or strong) correlation in a black-box manner have attracted much attention. In these theories, a spin projection operator is used to recover the spin symmetry of a broken-symmetry Slater determinant. Very recently, Pons Viver proposed the practical and exact implementation of Löwdin's spin projection operator [Int. J. Quantum Chem. **119**, e25770 (2019)]. In the present study, we attempt to supply mathematical proofs to Pons Viver's proposals and show a condition for establishing Pons Viver's implementation. Moreover, we explicitly derive the (spin projected) extended Hartree–Fock (EHF) equations on the basis of the model of common orbitals (i.e., closed-shell orbitals used in the restricted open-shell Hartree–Fock (ROHF) method), which was combined by Pons Viver with the EHF method.

## 1. Introduction

Recently, in the field of molecular electronic structure theory, Scuseria and co-workers has developed the projected Hartree–Fock (PHF) method in order to efficiently treat static (or strong) correlation in a black-box manner [1–4]. In the PHF method, to deliberatively break spin symmetry of a single Slater determinant and then variationally restore it, a spin projection operator such as Löwdin's spin projection operator [5] is employed [1]. Concretely, the projection operator is expressed as an integration of the spin-rotation operator (i.e., Percus-Rotenberg integration [6]) and it can be numerically evaluated with grid points [3,7]. As a result, one must determine the appropriate number of grid points, and then one must calculate the rotated Fock matrix at each grid point [7]. These results may give a negative image to the PHF method, but it has been reported that the PHF method has modest mean-field computational cost [1–3]. Moreover, according to Lestrange et al. [7], these inconveniences originating from the use of the grid points can be mitigated by using an efficient algorithm and Lebedev integration grids. On the other hand, unfortunately, the PHF method is not size consistent [1], but Scuseria and co-workers has attempted to reduce the size consistency errors while exploring the origin of the errors [8,9].

On the other hand, to obtain a wave function of the desired spin multiplicity from an unrestricted Hartree–Fock (UHF) wave function, Mayer et al. [10–12] already derived the (spin projected) extended Hartree–Fock (EHF) equations by using both Löwdin's spin projection operator and generalized Brillouin theorem [13–15] in the 1970s. It has been known that for small molecules the EHF method yields good results [12], but its computational cost is high due to its quite complicated equations [16]. Moreover, the EHF method is also not size consistent [16–18] and another serious shortcoming of the EHF method is that for large systems the EHF energy per particle reduces to that of the corresponding UHF wave function [18]. Conversely, complete equations for a second-order EHF orbital optimization scheme and wave function stability conditions, based on a second-order expansion of the EHF energy expression, have been published [19]. Moreover, there is an approach which provides almost EHF quality results at the computational expense of a UHF calculation, and it is called the half-projected HF approach [20,21]. Incidentally, the grid points of the PHF method are not employed in the EHF method, because Löwdin's spin projection operator is analytically evaluated in the EHF method [5,12].

In contrast, very recently, Pons Viver [22] proposed the practical and exact implementation of Löwdin's spin projection operator (i.e., the EHF method). The characteristic of Pons Viver's implementation is the practical use of the expectation value of Löwdin's spin projection operator. He denoted the expectation value as $\mathcal{G}(z)$, provided the practical and compact form of $\mathcal{G}(z)$, and showed that the total electronic energy of the EHF method is expressed in terms of $\mathcal{G}(z)$ [22]. Incidentally, Mayer et al. [10] also introduced the quantities $A_b^a$ corresponding to $\mathcal{G}(z)$ in order to express the EHF energy. Another substantial idea of Pons Viver may be the application of the model of common orbitals (CO) (or closed-shell orbitals) to the EHF method [22], although usually the CO model is adopted in the restricted open-shell Hartree–Fock (ROHF) method [23,24]. In fact, Pons Viver demonstrated that the CO model greatly simplifies the expression of the EHF energy [22]. (In the present study, we do not refer to the configuration interaction (CI) part of Pons Viver's paper [22].)

Although the spin-rotation operator is employed in the PHF method [1–4,7,8], we note that rotations of the spin quantization axis should be treated not classically but quantum mechanically [25–27]. Indeed, the so-called Pederson–Khanna (PK) method [28], which calculates zero-field splitting (ZFS) tensors at the density functional theory (DFT) level, was formulated on the basis of the classical rotations of the spin quantization axis, but later it has been pointed out that especially for triplet molecules the PK method systematically underestimates ZFS values by a factor of two [29]. According to Schmitt et al. [25], the commutation relation of the components of the spin operator is likely to be incorrect in the PK method (i.e., in the classical rotations). Incidentally, Neese [30] proposed the ZFS formulation based on the coupled-perturbed spin-orbit coupling (CP-SOC) equations and resolved the underestimations of the PK method. However, Schmitt et al. [25] and Yoshizawa et al. [26,27] later modified the CP-SOC-based ZFS formulation.

In the present study, first we attempt to prove that the rule of thumb used in Pons Viver's implementation [22] can mathematically be derived, and we show a condition for establishing the implementation. Next, following Pons Viver's proposals [22], we derive the EHF energy without using the CO model, while we explicitly derive the EHF energy and EHF equations on the basis of the CO model.

## 2. Expectation value of the spin projection operator

In the EHF method [5,12], the total electronic energy $E$ is defined by

$$E = \frac{\langle \Psi | \hat{\mathcal{P}}_S^\dagger \hat{H} \hat{\mathcal{P}}_S | \Psi \rangle}{\langle \Psi | \hat{\mathcal{P}}_S^\dagger \hat{\mathcal{P}}_S | \Psi \rangle} = \frac{\langle \Psi | \hat{H} \hat{\mathcal{P}}_S | \Psi \rangle}{\langle \Psi | \hat{\mathcal{P}}_S | \Psi \rangle}. \tag{1}$$

Here, $\hat{\mathcal{P}}_S$ is Löwdin's spin projection operator and is defined by

$$\hat{\mathcal{P}}_S = \prod_{\substack{l=-S \\ (l \neq S)}}^{S} \frac{\hat{S}^2 - l(l+1)}{S(S+1) - l(l+1)}, \tag{2}$$

where $\hat{\mathbf{S}}$ is the total spin operator, $S$ is the total spin quantum number, $S(S+1)$ is the eigenvalue of the operator $\hat{S}^2$, and $\hat{\mathcal{P}}_S^\dagger = \hat{\mathcal{P}}_S = \hat{\mathcal{P}}_S^2$ [5,12]. The wave function $|\Psi\rangle$ is an UHF Slater determinant, and we assume that the UHF Slater determinant $|\Psi\rangle$ consists of $\mu$ spin-orbitals with the spin function $|\alpha\rangle$ and $\nu$ spin-orbitals with the spin function $|\beta\rangle$ ($\mu \geq \nu$). Also, $\hat{H}$ denotes a non-relativistic or scalar relativistic Hamiltonian, and then $\hat{\mathcal{P}}_S \hat{H} = \hat{H} \hat{\mathcal{P}}_S$ [5,12].

The action of the projection operator $\hat{\mathcal{P}}_S$ on the UHF Slater determinant $|\Psi\rangle$ can be written as a linear combination of configuration state functions $|T_k\rangle$ [5,31,32]:

$$\hat{\mathcal{P}}_S | \Psi \rangle = \sum_{k=0}^{\nu} C_k(S,M,N) | T_k \rangle. \tag{3}$$

Here, $C_k(S,M,N)$ are Sanibel coefficients [31,33], $M = (\mu - \nu)/2$, and $N = (\mu + \nu)/2$ [22]. Hereafter, we often adopt a shorthand notation $C_k \coloneqq C_k(S,M,N)$. The function $|T_k\rangle$ is the sum of all the $n_k$ Slater determinants obtained from $|\Psi\rangle$ by replacing $k$ spins $\alpha$ by $\beta$ and, simultaneously, $k$ spins $\beta$ by $\alpha$:

$$|T_k\rangle = |T_k^{(1)}\rangle + |T_k^{(2)}\rangle + \cdots + |T_k^{(n_k)}\rangle, \tag{4}$$

where $n_k = {}_\mu C_k \cdot {}_\nu C_k$ [5,12,32]. For example, the UHF Slater determinant $|\Psi\rangle$ is given by [22]

$$|\Psi\rangle = |T_0\rangle = \left| \phi_1^\alpha \cdots \phi_{i_1-1}^\alpha \phi_{i_1}^\alpha \phi_{i_1+1}^\alpha \cdots \phi_{i_k-1}^\alpha \phi_{i_k}^\alpha \phi_{i_k+1}^\alpha \cdots \phi_\mu^\alpha \ \theta_1^\beta \cdots \theta_{j_1-1}^\beta \theta_{j_1}^\beta \theta_{j_1+1}^\beta \cdots \theta_{j_k-1}^\beta \theta_{j_k}^\beta \theta_{j_k+1}^\beta \cdots \theta_\nu^\beta \right|. \tag{5}$$

Here, $\phi_l^\alpha = \phi_l |\alpha\rangle$ where $\phi_l$ is the spatial part of a spin-orbital $\phi_l^\alpha$ and $\theta_l^\beta = \theta_l |\beta\rangle$ where $\theta_l$ is the spatial part of a spin-orbital $\theta_l^\beta$. Then, the function $|T_k\rangle$ is given by [22]

$$|T_k\rangle = \sum_{\substack{1 \leq i_1 < i_2 < \cdots < i_k \leq \mu \\ 1 \leq j_1 < j_2 < \cdots < j_k \leq \nu}} \left| \phi_1^\alpha \cdots \phi_{i_1-1}^\alpha \phi_{j_1}^\beta \phi_{i_1+1}^\alpha \cdots \phi_{i_k-1}^\alpha \phi_{j_k}^\beta \phi_{i_k+1}^\alpha \cdots \phi_\mu^\alpha \ \theta_1^\beta \cdots \theta_{j_1-1}^\beta \theta_{i_1}^\alpha \theta_{j_1+1}^\beta \cdots \theta_{j_k-1}^\beta \theta_{i_k}^\alpha \theta_{j_k+1}^\beta \cdots \theta_\nu^\beta \right|. \tag{6}$$

For later convenience, by using the property of determinants, we can rewrite Eq. (6) as [22]

$$|T_k\rangle = \sum_{\substack{1 \leq i_1 < i_2 < \cdots < i_k \leq \mu \\ 1 \leq j_1 < j_2 < \cdots < j_k \leq \nu}} \left| \phi_1^\alpha \cdots \phi_{i_1-1}^\alpha \theta_{j_1}^\alpha \phi_{i_1+1}^\alpha \cdots \phi_{i_k-1}^\alpha \theta_{j_k}^\alpha \phi_{i_k+1}^\alpha \cdots \phi_\mu^\alpha \ \theta_1^\beta \cdots \theta_{j_1-1}^\beta \phi_{i_1}^\beta \theta_{j_1+1}^\beta \cdots \theta_{j_k-1}^\beta \phi_{i_k}^\beta \theta_{j_k+1}^\beta \cdots \theta_\nu^\beta \right| \cdot (-1)^k. \tag{7}$$

Hence, the expectation value of the projection operator $\hat{\mathcal{P}}_S$ can be written as

$$\langle \Psi | \hat{\mathcal{P}}_S | \Psi \rangle = \sum_{k=0}^\nu C_k \langle T_0 | T_k \rangle$$

$$= \sum_{k=0}^\nu \sum_{\substack{1 \leq i_1 < i_2 < \cdots < i_k \leq \mu \\ 1 \leq j_1 < j_2 < \cdots < j_k \leq \nu}} (-1)^k C_k \langle T_0 | T_k^{i_k j_k} \rangle$$

$$= \sum_{k=0}^\nu (-1)^k C_k \sum_{\substack{1 \leq i_1 < i_2 < \cdots < i_k \leq \mu \\ 1 \leq j_1 < j_2 < \cdots < j_k \leq \nu}} \begin{vmatrix} \langle \phi_{i_1} | \theta_{j_1} \rangle & \cdots & \langle \phi_{i_1} | \theta_{j_k} \rangle \\ \vdots & \ddots & \vdots \\ \langle \phi_{i_k} | \theta_{j_1} \rangle & \cdots & \langle \phi_{i_k} | \theta_{j_k} \rangle \end{vmatrix} \cdot \begin{vmatrix} \langle \theta_{j_1} | \phi_{i_1} \rangle & \cdots & \langle \theta_{j_1} | \phi_{i_k} \rangle \\ \vdots & \ddots & \vdots \\ \langle \theta_{j_k} | \phi_{i_1} \rangle & \cdots & \langle \theta_{j_k} | \phi_{i_k} \rangle \end{vmatrix}, \tag{8}$$

where $|T_k^{i_k j_k}\rangle$ is the determinant used in Eq. (7):

$$\left| T_k^{i_k j_k} \right\rangle := \left| \phi_1^\alpha \cdots \phi_{i_1-1}^\alpha \theta_{j_1}^\alpha \phi_{i_1+1}^\alpha \cdots \phi_{i_k-1}^\alpha \theta_{j_k}^\alpha \phi_{i_k+1}^\alpha \cdots \phi_\mu^\alpha \ \theta_1^\beta \cdots \theta_{j_1-1}^\beta \phi_{i_1}^\beta \theta_{j_1+1}^\beta \cdots \theta_{j_k-1}^\beta \phi_{i_k}^\beta \theta_{j_k+1}^\beta \cdots \theta_\nu^\beta \right|,$$

$$|T_k\rangle = \sum_{\substack{1 \leq i_1 < i_2 < \cdots < i_k \leq \mu \\ 1 \leq j_1 < j_2 < \cdots < j_k \leq \nu}} (-1)^k \left| T_k^{i_k j_k} \right\rangle. \tag{9}$$

To understand the calculation of $\langle T_0 | T_k^{i_k j_k} \rangle$ in Eq. (8), we consider a simple example:

$$|T_0\rangle = \left| \phi_1^\alpha \phi_2^\alpha \phi_3^\alpha \phi_4^\alpha \phi_5^\alpha \phi_6^\alpha \ \theta_1^\beta \theta_2^\beta \theta_3^\beta \theta_4^\beta \right|,$$

$$\left| T_k^{i_k j_k} \right\rangle = \left| \phi_1^\alpha \phi_2^\alpha \phi_3^\alpha \theta_2^\alpha \theta_3^\alpha \phi_6^\alpha \ \theta_1^\beta \phi_4^\beta \phi_5^\beta \theta_4^\beta \right|. \tag{10}$$

Then, we have

$$\langle T_0 | T_k^{i_k j_k} \rangle =$$

$$\begin{vmatrix} \langle \phi_1^\alpha | \phi_1^\alpha \rangle & \langle \phi_1^\alpha | \phi_2^\alpha \rangle & \langle \phi_1^\alpha | \phi_3^\alpha \rangle & \langle \phi_1^\alpha | \theta_2^\alpha \rangle & \langle \phi_1^\alpha | \theta_3^\alpha \rangle & \langle \phi_1^\alpha | \phi_6^\alpha \rangle & \langle \phi_1^\alpha | \theta_1^\beta \rangle & \langle \phi_1^\alpha | \phi_4^\beta \rangle & \langle \phi_1^\alpha | \phi_5^\beta \rangle & \langle \phi_1^\alpha | \theta_4^\beta \rangle \\ \langle \phi_2^\alpha | \phi_1^\alpha \rangle & \langle \phi_2^\alpha | \phi_2^\alpha \rangle & \langle \phi_2^\alpha | \phi_3^\alpha \rangle & \langle \phi_2^\alpha | \theta_2^\alpha \rangle & \langle \phi_2^\alpha | \theta_3^\alpha \rangle & \langle \phi_2^\alpha | \phi_6^\alpha \rangle & \langle \phi_2^\alpha | \theta_1^\beta \rangle & \langle \phi_2^\alpha | \phi_4^\beta \rangle & \langle \phi_2^\alpha | \phi_5^\beta \rangle & \langle \phi_2^\alpha | \theta_4^\beta \rangle \\ \langle \phi_3^\alpha | \phi_1^\alpha \rangle & \langle \phi_3^\alpha | \phi_2^\alpha \rangle & \langle \phi_3^\alpha | \phi_3^\alpha \rangle & \langle \phi_3^\alpha | \theta_2^\alpha \rangle & \langle \phi_3^\alpha | \theta_3^\alpha \rangle & \langle \phi_3^\alpha | \phi_6^\alpha \rangle & \langle \phi_3^\alpha | \theta_1^\beta \rangle & \langle \phi_3^\alpha | \phi_4^\beta \rangle & \langle \phi_3^\alpha | \phi_5^\beta \rangle & \langle \phi_3^\alpha | \theta_4^\beta \rangle \\ \langle \phi_4^\alpha | \phi_1^\alpha \rangle & \langle \phi_4^\alpha | \phi_2^\alpha \rangle & \langle \phi_4^\alpha | \phi_3^\alpha \rangle & \langle \phi_4^\alpha | \theta_2^\alpha \rangle & \langle \phi_4^\alpha | \theta_3^\alpha \rangle & \langle \phi_4^\alpha | \phi_6^\alpha \rangle & \langle \phi_4^\alpha | \theta_1^\beta \rangle & \langle \phi_4^\alpha | \phi_4^\beta \rangle & \langle \phi_4^\alpha | \phi_5^\beta \rangle & \langle \phi_4^\alpha | \theta_4^\beta \rangle \\ \langle \phi_5^\alpha | \phi_1^\alpha \rangle & \langle \phi_5^\alpha | \phi_2^\alpha \rangle & \langle \phi_5^\alpha | \phi_3^\alpha \rangle & \langle \phi_5^\alpha | \theta_2^\alpha \rangle & \langle \phi_5^\alpha | \theta_3^\alpha \rangle & \langle \phi_5^\alpha | \phi_6^\alpha \rangle & \langle \phi_5^\alpha | \theta_1^\beta \rangle & \langle \phi_5^\alpha | \phi_4^\beta \rangle & \langle \phi_5^\alpha | \phi_5^\beta \rangle & \langle \phi_5^\alpha | \theta_4^\beta \rangle \\ \langle \phi_6^\alpha | \phi_1^\alpha \rangle & \langle \phi_6^\alpha | \phi_2^\alpha \rangle & \langle \phi_6^\alpha | \phi_3^\alpha \rangle & \langle \phi_6^\alpha | \theta_2^\alpha \rangle & \langle \phi_6^\alpha | \theta_3^\alpha \rangle & \langle \phi_6^\alpha | \phi_6^\alpha \rangle & \langle \phi_6^\alpha | \theta_1^\beta \rangle & \langle \phi_6^\alpha | \phi_4^\beta \rangle & \langle \phi_6^\alpha | \phi_5^\beta \rangle & \langle \phi_6^\alpha | \theta_4^\beta \rangle \\ \langle \theta_1^\beta | \phi_1^\alpha \rangle & \langle \theta_1^\beta | \phi_2^\alpha \rangle & \langle \theta_1^\beta | \phi_3^\alpha \rangle & \langle \theta_1^\beta | \theta_2^\alpha \rangle & \langle \theta_1^\beta | \theta_3^\alpha \rangle & \langle \theta_1^\beta | \phi_6^\alpha \rangle & \langle \theta_1^\beta | \theta_1^\beta \rangle & \langle \theta_1^\beta | \phi_4^\beta \rangle & \langle \theta_1^\beta | \phi_5^\beta \rangle & \langle \theta_1^\beta | \theta_4^\beta \rangle \\ \langle \theta_2^\beta | \phi_1^\alpha \rangle & \langle \theta_2^\beta | \phi_2^\alpha \rangle & \langle \theta_2^\beta | \phi_3^\alpha \rangle & \langle \theta_2^\beta | \theta_2^\alpha \rangle & \langle \theta_2^\beta | \theta_3^\alpha \rangle & \langle \theta_2^\beta | \phi_6^\alpha \rangle & \langle \theta_2^\beta | \theta_1^\beta \rangle & \langle \theta_2^\beta | \phi_4^\beta \rangle & \langle \theta_2^\beta | \phi_5^\beta \rangle & \langle \theta_2^\beta | \theta_4^\beta \rangle \\ \langle \theta_3^\beta | \phi_1^\alpha \rangle & \langle \theta_3^\beta | \phi_2^\alpha \rangle & \langle \theta_3^\beta | \phi_3^\alpha \rangle & \langle \theta_3^\beta | \theta_2^\alpha \rangle & \langle \theta_3^\beta | \theta_3^\alpha \rangle & \langle \theta_3^\beta | \phi_6^\alpha \rangle & \langle \theta_3^\beta | \theta_1^\beta \rangle & \langle \theta_3^\beta | \phi_4^\beta \rangle & \langle \theta_3^\beta | \phi_5^\beta \rangle & \langle \theta_3^\beta | \theta_4^\beta \rangle \\ \langle \theta_4^\beta | \phi_1^\alpha \rangle & \langle \theta_4^\beta | \phi_2^\alpha \rangle & \langle \theta_4^\beta | \phi_3^\alpha \rangle & \langle \theta_4^\beta | \theta_2^\alpha \rangle & \langle \theta_4^\beta | \theta_3^\alpha \rangle & \langle \theta_4^\beta | \phi_6^\alpha \rangle & \langle \theta_4^\beta | \theta_1^\beta \rangle & \langle \theta_4^\beta | \phi_4^\beta \rangle & \langle \theta_4^\beta | \phi_5^\beta \rangle & \langle \theta_4^\beta | \theta_4^\beta \rangle \end{vmatrix}$$

$$= \begin{vmatrix} 1 & 0 & 0 & \langle \phi_1|\theta_2\rangle & \langle \phi_1|\theta_3\rangle & 0 & 0 & 0 & 0 & 0 \\ 0 & 1 & 0 & \langle \phi_2|\theta_2\rangle & \langle \phi_2|\theta_3\rangle & 0 & 0 & 0 & 0 & 0 \\ 0 & 0 & 1 & \langle \phi_3|\theta_2\rangle & \langle \phi_3|\theta_3\rangle & 0 & 0 & 0 & 0 & 0 \\ 0 & 0 & 0 & \langle \phi_4|\theta_2\rangle & \langle \phi_4|\theta_3\rangle & 0 & 0 & 0 & 0 & 0 \\ 0 & 0 & 0 & \langle \phi_5|\theta_2\rangle & \langle \phi_5|\theta_3\rangle & 0 & 0 & 0 & 0 & 0 \\ 0 & 0 & 0 & \langle \phi_6|\theta_2\rangle & \langle \phi_6|\theta_3\rangle & 1 & 0 & 0 & 0 & 0 \\ 0 & 0 & 0 & 0 & 0 & 0 & 1 & \langle \theta_1|\phi_4\rangle & \langle \theta_1|\phi_5\rangle & 0 \\ 0 & 0 & 0 & 0 & 0 & 0 & 0 & \langle \theta_2|\phi_4\rangle & \langle \theta_2|\phi_5\rangle & 0 \\ 0 & 0 & 0 & 0 & 0 & 0 & 0 & \langle \theta_3|\phi_4\rangle & \langle \theta_3|\phi_5\rangle & 0 \\ 0 & 0 & 0 & 0 & 0 & 0 & 0 & \langle \theta_4|\phi_4\rangle & \langle \theta_4|\phi_5\rangle & 1 \end{vmatrix}$$

$$= \begin{vmatrix} 1 & 0 & 0 & \langle \phi_1|\theta_2\rangle & \langle \phi_1|\theta_3\rangle & 0 \\ 0 & 1 & 0 & \langle \phi_2|\theta_2\rangle & \langle \phi_2|\theta_3\rangle & 0 \\ 0 & 0 & 1 & \langle \phi_3|\theta_2\rangle & \langle \phi_3|\theta_3\rangle & 0 \\ 0 & 0 & 0 & \langle \phi_4|\theta_2\rangle & \langle \phi_4|\theta_3\rangle & 0 \\ 0 & 0 & 0 & \langle \phi_5|\theta_2\rangle & \langle \phi_5|\theta_3\rangle & 0 \\ 0 & 0 & 0 & \langle \phi_6|\theta_2\rangle & \langle \phi_6|\theta_3\rangle & 1 \end{vmatrix} \cdot \begin{vmatrix} 1 & \langle \theta_1|\phi_4\rangle & \langle \theta_1|\phi_5\rangle & 0 \\ 0 & \langle \theta_2|\phi_4\rangle & \langle \theta_2|\phi_5\rangle & 0 \\ 0 & \langle \theta_3|\phi_4\rangle & \langle \theta_3|\phi_5\rangle & 0 \\ 0 & \langle \theta_4|\phi_4\rangle & \langle \theta_4|\phi_5\rangle & 1 \end{vmatrix} = \begin{vmatrix} \langle \phi_4|\theta_2\rangle & \langle \phi_4|\theta_3\rangle \\ \langle \phi_5|\theta_2\rangle & \langle \phi_5|\theta_3\rangle \end{vmatrix} \cdot \begin{vmatrix} \langle \theta_2|\phi_4\rangle & \langle \theta_2|\phi_5\rangle \\ \langle \theta_3|\phi_4\rangle & \langle \theta_3|\phi_5\rangle \end{vmatrix},$$

(11)

where some properties of determinants have been employed [34].

On the other hand, Pons Viver [22] proved that the expectation value $\langle \Psi | \hat{\mathcal{P}}_S | \Psi \rangle$ of Eq. (8) can be expressed by the determinant of the following matrix,

$$\mathbf{G}(z) = \begin{pmatrix} \mathbf{I}_{\mu \times \mu} & \mathbf{C} \\ \mathbf{C}^\dagger & z \mathbf{I}_{\nu \times \nu} \end{pmatrix}, \tag{12}$$

where $z$ is a formal variable associated with the Sanibel coefficients $C_k(S, M, N)$, $\mathbf{C}$ ($= (c_{ij})$) is the $\mu \times \nu$ matrix of overlaps between the spatial orbitals,

$$\mathbf{C} = \begin{pmatrix} \langle \phi_1 | \theta_1 \rangle & \cdots & \langle \phi_1 | \theta_\nu \rangle \\ \vdots & \ddots & \vdots \\ \langle \phi_\mu | \theta_1 \rangle & \cdots & \langle \phi_\mu | \theta_\nu \rangle \end{pmatrix}, \tag{13}$$

and $\mathbf{I}_{\mu\times\mu}$ and $\mathbf{I}_{\nu\times\nu}$ are $\mu\times\mu$ and $\nu\times\nu$ unit matrices, respectively. Here, we provide the proof which differs from Pons Viver's proof [22] as follows. The determinant of $\mathbf{G}(z)$ becomes

$$\left|\mathbf{G}(z)\right| = \begin{vmatrix} \mathbf{I}_{\mu\times\mu} & \mathbf{C} \\ \mathbf{C}^{\dagger} & z\mathbf{I}_{\nu\times\nu} \end{vmatrix} = \left|z\cdot\mathbf{I}_{\nu\times\nu} - \mathbf{C}^{\dagger}\mathbf{C}\right|, \qquad (14)$$

where we have used the following property of determinants:

$$\det\begin{pmatrix} \mathbf{A} & \mathbf{C} \\ \mathbf{B} & \mathbf{D} \end{pmatrix} = \det(\mathbf{A})\det(\mathbf{D} - \mathbf{B}\mathbf{A}^{-1}\mathbf{C}). \qquad (15)$$

Because the right hand side of Eq. (14) is the characteristic polynomial of the $\nu\times\nu$ matrix $\mathbf{M} \coloneqq \mathbf{C}^{\dagger}\mathbf{C} = (m_{ij})$, we have

$$\left|z\cdot\mathbf{I}_{\nu\times\nu} - \mathbf{C}^{\dagger}\mathbf{C}\right| = \sum_{k=0}^{\nu} d_k z^{\nu-k}, \qquad (16)$$

where the coefficients $d_k$ are generally given by

$$d_k = (-1)^k \sum_{1\le j_1<\cdots<j_k\le\nu} \begin{vmatrix} m_{j_1 j_1} & m_{j_1 j_2} & \cdots & m_{j_1 j_k} \\ m_{j_2 j_1} & m_{j_2 j_2} & \cdots & m_{j_2 j_k} \\ \vdots & \vdots & \ddots & \vdots \\ m_{j_k j_1} & m_{j_k j_2} & \cdots & m_{j_k j_k} \end{vmatrix}. \qquad (17)$$

Then, we apply the generalized Cauchy–Binet theorem [35] to the determinants in Eq. (17):

$$\begin{vmatrix} m_{j_1 j_1} & m_{j_1 j_2} & \cdots & m_{j_1 j_k} \\ m_{j_2 j_1} & m_{j_2 j_2} & \cdots & m_{j_2 j_k} \\ \vdots & \vdots & \ddots & \vdots \\ m_{j_k j_1} & m_{j_k j_2} & \cdots & m_{j_k j_k} \end{vmatrix} = \begin{vmatrix} \mathbf{C}^{*}_{\mu\times j_1}\mathbf{C}_{\mu\times j_1} & \cdots & \mathbf{C}^{*}_{\mu\times j_1}\mathbf{C}_{\mu\times j_k} \\ \vdots & \ddots & \vdots \\ \mathbf{C}^{*}_{\mu\times j_k}\mathbf{C}_{\mu\times j_1} & \cdots & \mathbf{C}^{*}_{\mu\times j_k}\mathbf{C}_{\mu\times j_k} \end{vmatrix}$$

$$= \sum_{1\le i_1<\cdots<i_k\le\mu} \begin{vmatrix} c^{*}_{i_1 j_1} & \cdots & c^{*}_{i_k j_1} \\ \vdots & \ddots & \vdots \\ c^{*}_{i_1 j_k} & \cdots & c^{*}_{i_k j_k} \end{vmatrix} \cdot \begin{vmatrix} c_{i_1 j_1} & \cdots & c_{i_1 j_k} \\ \vdots & \ddots & \vdots \\ c_{i_k j_1} & \cdots & c_{i_k j_k} \end{vmatrix}$$

$$= \sum_{1\le i_1<\cdots<i_k\le\mu} \begin{vmatrix} \langle\theta_{j_1}|\phi_{i_1}\rangle & \cdots & \langle\theta_{j_1}|\phi_{i_k}\rangle \\ \vdots & \ddots & \vdots \\ \langle\theta_{j_k}|\phi_{i_1}\rangle & \cdots & \langle\theta_{j_k}|\phi_{i_k}\rangle \end{vmatrix} \cdot \begin{vmatrix} \langle\phi_{i_1}|\theta_{j_1}\rangle & \cdots & \langle\phi_{i_1}|\theta_{j_k}\rangle \\ \vdots & \ddots & \vdots \\ \langle\phi_{i_k}|\theta_{j_1}\rangle & \cdots & \langle\phi_{i_k}|\theta_{j_k}\rangle \end{vmatrix}. \qquad (18)$$

Thus, substitutions of Eq. (18) into Eq. (17), Eq. (17) into Eq. (16), and Eq. (16) into Eq. (14) yield

$$\left|\mathbf{G}(z)\right| = \sum_{k=0}^{\nu} (-1)^k z^{\nu-k} \sum_{\substack{1\le i_1<\cdots<i_k\le\mu \\ 1\le j_1<\cdots<j_k\le\nu}} \begin{vmatrix} \langle\phi_{i_1}|\theta_{j_1}\rangle & \cdots & \langle\phi_{i_1}|\theta_{j_k}\rangle \\ \vdots & \ddots & \vdots \\ \langle\phi_{i_k}|\theta_{j_1}\rangle & \cdots & \langle\phi_{i_k}|\theta_{j_k}\rangle \end{vmatrix} \cdot \begin{vmatrix} \langle\theta_{j_1}|\phi_{i_1}\rangle & \cdots & \langle\theta_{j_1}|\phi_{i_k}\rangle \\ \vdots & \ddots & \vdots \\ \langle\theta_{j_k}|\phi_{i_1}\rangle & \cdots & \langle\theta_{j_k}|\phi_{i_k}\rangle \end{vmatrix}. \qquad (19)$$

Finally, when letting $z^{\nu-k}$ be associated with $C_k(S,M,N)$, we obtain $\left|\mathbf{G}(z)\right| = \langle\Psi|\hat{\mathcal{P}}_S|\Psi\rangle$ (please compare Eq. (8) and Eq. (19)). Following Pons Viver's notation [22], we define $\mathcal{G}(z)$ as

$$\mathcal{G}(z) := \det\left(z \cdot \mathbf{I}_{\nu\times\nu} - \mathbf{C}^\dagger \mathbf{C}\right) = |\mathbf{G}(z)|. \tag{20}$$

## 3. Non-orthogonal matrix elements of one-electron operators

To express the EHF energy of Eq. (1) in terms of $\mathcal{G}(z)$, we consider the following expectation value $\mathcal{F}(C_k)$ of one-electron operators:

$$\mathcal{F}(C_k) = \langle \Psi | \sum_{i=1}^{n} \hat{h}(\mathbf{r}_i) | \hat{\mathcal{P}}_S \Psi \rangle, \tag{21}$$

where $n\ (= \mu + \nu)$ is the number of electrons and $\hat{h}(\mathbf{r}_i)$ is an one-electron operator for electron $i$. Then, substituting Eqs. (3) and (9) into Eq. (21) gives

$$\begin{aligned}
\mathcal{F}(C_k) &= \sum_{k=0}^{\nu} C_k \langle T_0 | \sum_{i=1}^{n} h(\mathbf{r}_i) | T_k \rangle \\
&= \sum_{k=0}^{\nu} (-1)^k C_k \sum_{\substack{1 \le i_1 < i_2 < \cdots < i_k \le \mu \\ 1 \le j_1 < j_2 < \cdots < j_k \le \nu}} \langle T_0 | \sum_{i=1}^{n} h(\mathbf{r}_i) | T_k^{i_k j_k} \rangle.
\end{aligned} \tag{22}$$

Generally, the one-electron operator matrix elements between non-orthogonal Slater determinants can be written as [36,37]

$$\begin{aligned}
\langle \Phi_F | \sum_{i=1}^{n} \hat{h}(\mathbf{r}_i) | \Phi_G \rangle &= \sum_{i=1}^{n} \sum_{j=1}^{n} \langle f_i | \hat{h} | g_j \rangle \cdot (-1)^{i+j} \left| \mathbf{S}^{(i,j)} \right| \\
&= \sum_{i=1}^{n} \sum_{j=1}^{n} \langle f_i | \hat{h} | g_j \rangle \cdot \mathrm{adj}(\mathbf{S})_{ji} \\
&= \mathrm{tr}\left[ \mathbf{h} \cdot \mathrm{adj}(\mathbf{S}) \right],
\end{aligned} \tag{23}$$

where

$$\mathbf{S} := \begin{pmatrix} \langle f_1 | g_1 \rangle & \langle f_1 | g_2 \rangle & \cdots & \langle f_1 | g_n \rangle \\ \langle f_2 | g_1 \rangle & \langle f_2 | g_2 \rangle & \cdots & \langle f_2 | g_n \rangle \\ \vdots & \vdots & & \vdots \\ \langle f_n | g_1 \rangle & \langle f_n | g_2 \rangle & \cdots & \langle f_n | g_n \rangle \end{pmatrix}, \tag{24}$$

$f_i$ is the $i$th spin-orbital in the Slater determinant $\Phi_F$, $g_j$ is the $j$th spin-orbital in the Slater determinant $\Phi_G$, $\left|\mathbf{S}^{(i,j)}\right|$ is the determinant obtained from $|\mathbf{S}|$ by deleting the $i$th row and the $j$th column, and $\mathrm{adj}(\mathbf{S})$ is the adjugate matrix of the overlap matrix $\mathbf{S}$. Thus, by applying Eq. (23) to Eq. (22), we obtain

$$\mathcal{F}(C_k) = \sum_{k=0}^{\nu} (-1)^k C_k \sum_{\substack{1 \le i_1 < i_2 < \cdots < i_k \le \mu \\ 1 \le j_1 < j_2 < \cdots < j_k \le \nu}} \mathrm{tr}\left[ \mathbf{h}^{i_k j_k} \cdot \mathrm{adj}\left(\mathbf{S}^{i_k j_k}\right) \right], \tag{25}$$

where

$$\mathbf{h}^{i_k j_k} = \begin{pmatrix} \mathbf{h}^{\alpha\alpha}_{\leq j_k} & 0 \\ 0 & \mathbf{h}^{\beta\beta}_{\leq i_k} \end{pmatrix}, \qquad \mathbf{S}^{i_k j_k} = \begin{pmatrix} \mathbf{II}^{\alpha\alpha}_{\leq j_k} & 0 \\ 0 & \mathbf{II}^{\beta\beta}_{\leq i_k} \end{pmatrix},$$

$\mathbf{h}^{\alpha\alpha}_{\leq j_k} =$

$$\begin{pmatrix}
\langle \phi_1|\hat{h}|\phi_1\rangle & \cdots & \langle \phi_1|\hat{h}|\phi_{i_1-1}\rangle & \langle \phi_1|\hat{h}|\theta_{j_1}\rangle & \langle \phi_1|\hat{h}|\phi_{i_1+1}\rangle & \cdots & \langle \phi_1|\hat{h}|\phi_{i_k-1}\rangle & \langle \phi_1|\hat{h}|\theta_{j_k}\rangle & \langle \phi_1|\hat{h}|\phi_{i_k+1}\rangle & \cdots & \langle \phi_1|\hat{h}|\phi_\mu\rangle \\
\vdots & \ddots & \vdots & \vdots & & & & & & & \vdots \\
\langle \phi_{i_1-1}|\hat{h}|\phi_1\rangle & \cdots & \langle \phi_{i_1-1}|\hat{h}|\phi_{i_1-1}\rangle & \langle \phi_{i_1-1}|\hat{h}|\theta_{j_1}\rangle & & & & & & & \langle \phi_{i_1-1}|\hat{h}|\phi_\mu\rangle \\
\langle \phi_{i_1}|\hat{h}|\phi_1\rangle & \cdots & \langle \phi_{i_1}|\hat{h}|\phi_{i_1-1}\rangle & \langle \phi_{i_1}|\hat{h}|\theta_{j_1}\rangle & & & & & & & \langle \phi_{i_1}|\hat{h}|\phi_\mu\rangle \\
\langle \phi_{i_1+1}|\hat{h}|\phi_1\rangle & \cdots & \langle \phi_{i_1+1}|\hat{h}|\phi_{i_1-1}\rangle & \langle \phi_{i_1+1}|\hat{h}|\theta_{j_1}\rangle & \ddots & & & & & & \langle \phi_{i_1+1}|\hat{h}|\phi_\mu\rangle \\
\vdots & \vdots & \vdots & \vdots & & \ddots & & & & & \vdots \\
\langle \phi_{i_k-1}|\hat{h}|\phi_1\rangle & \cdots & \langle \phi_{i_k-1}|\hat{h}|\phi_{i_1-1}\rangle & \langle \phi_{i_k-1}|\hat{h}|\theta_{j_1}\rangle & & & \ddots & & & & \langle \phi_{i_k-1}|\hat{h}|\phi_\mu\rangle \\
\langle \phi_{i_k}|\hat{h}|\phi_1\rangle & \cdots & \langle \phi_{i_k}|\hat{h}|\phi_{i_1-1}\rangle & \langle \phi_{i_k}|\hat{h}|\theta_{j_1}\rangle & & & & \ddots & & & \langle \phi_{i_k}|\hat{h}|\phi_\mu\rangle \\
\langle \phi_{i_k+1}|\hat{h}|\phi_1\rangle & \cdots & \langle \phi_{i_k+1}|\hat{h}|\phi_{i_1-1}\rangle & \langle \phi_{i_k+1}|\hat{h}|\theta_{j_1}\rangle & & & & & \ddots & & \langle \phi_{i_k+1}|\hat{h}|\phi_\mu\rangle \\
\vdots & \vdots & \vdots & \vdots & & & & & & \ddots & \vdots \\
\langle \phi_\mu|\hat{h}|\phi_1\rangle & \cdots & \langle \phi_\mu|\hat{h}|\phi_{i_1-1}\rangle & \langle \phi_\mu|\hat{h}|\theta_{j_1}\rangle & \langle \phi_\mu|\hat{h}|\phi_{i_1+1}\rangle & \cdots & \langle \phi_\mu|\hat{h}|\phi_{i_k-1}\rangle & \langle \phi_\mu|\hat{h}|\theta_{j_k}\rangle & \langle \phi_\mu|\hat{h}|\phi_{i_k+1}\rangle & \cdots & \langle \phi_\mu|\hat{h}|\phi_\mu\rangle
\end{pmatrix},$$

$\mathbf{II}^{\alpha\alpha}_{\leq j_k} =$

$$\begin{pmatrix}
\langle \phi_1|\phi_1\rangle & \cdots & \langle \phi_1|\phi_{i_1-1}\rangle & \langle \phi_1|\theta_{j_1}\rangle & \langle \phi_1|\phi_{i_1+1}\rangle & \cdots & \langle \phi_1|\phi_{i_k-1}\rangle & \langle \phi_1|\theta_{j_k}\rangle & \langle \phi_1|\phi_{i_k+1}\rangle & \cdots & \langle \phi_1|\phi_\mu\rangle \\
\vdots & \ddots & \vdots & \vdots & & & & & & & \vdots \\
\langle \phi_{i_1-1}|\phi_1\rangle & \cdots & \langle \phi_{i_1-1}|\phi_{i_1-1}\rangle & \langle \phi_{i_1-1}|\theta_{j_1}\rangle & & & & & & & \langle \phi_{i_1-1}|\phi_\mu\rangle \\
\langle \phi_{i_1}|\phi_1\rangle & \cdots & \langle \phi_{i_1}|\phi_{i_1-1}\rangle & \langle \phi_{i_1}|\theta_{j_1}\rangle & & & & & & & \langle \phi_{i_1}|\phi_\mu\rangle \\
\langle \phi_{i_1+1}|\phi_1\rangle & \cdots & \langle \phi_{i_1+1}|\phi_{i_1-1}\rangle & \langle \phi_{i_1+1}|\theta_{j_1}\rangle & \ddots & & & & & & \langle \phi_{i_1+1}|\phi_\mu\rangle \\
\vdots & \vdots & \vdots & \vdots & & \ddots & & & & & \vdots \\
\langle \phi_{i_k-1}|\phi_1\rangle & \cdots & \langle \phi_{i_k-1}|\phi_{i_1-1}\rangle & \langle \phi_{i_k-1}|\theta_{j_1}\rangle & & & \ddots & & & & \langle \phi_{i_k-1}|\phi_\mu\rangle \\
\langle \phi_{i_k}|\phi_1\rangle & \cdots & \langle \phi_{i_k}|\phi_{i_1-1}\rangle & \langle \phi_{i_k}|\theta_{j_1}\rangle & & & & \ddots & & & \langle \phi_{i_k}|\phi_\mu\rangle \\
\langle \phi_{i_k+1}|\phi_1\rangle & \cdots & \langle \phi_{i_k+1}|\phi_{i_1-1}\rangle & \langle \phi_{i_k+1}|\theta_{j_1}\rangle & & & & & \ddots & & \langle \phi_{i_k+1}|\phi_\mu\rangle \\
\vdots & \vdots & \vdots & \vdots & & & & & & \ddots & \vdots \\
\langle \phi_\mu|\phi_1\rangle & \cdots & \langle \phi_\mu|\phi_{i_1-1}\rangle & \langle \phi_\mu|\theta_{j_1}\rangle & \langle \phi_\mu|\phi_{i_1+1}\rangle & \cdots & \langle \phi_\mu|\phi_{i_k-1}\rangle & \langle \phi_\mu|\theta_{j_k}\rangle & \langle \phi_\mu|\phi_{i_k+1}\rangle & \cdots & \langle \phi_\mu|\phi_\mu\rangle
\end{pmatrix}.$$

(26)

The zero matrices **0** in the matrices $\mathbf{h}^{i_k j_k}$ and $\mathbf{S}^{i_k j_k}$ are attributed to $\langle \alpha|\beta\rangle = 0$ or $\langle \beta|\alpha\rangle = 0$. When adopting the example of Eq. (10), by using $\langle \alpha|\alpha\rangle = 1$ and $\langle \phi_i|\phi_j\rangle = \delta_{ij}$, we can write the matrices $\mathbf{h}^{\alpha\alpha}_{\leq j_k}$ and $\mathbf{II}^{\alpha\alpha}_{\leq j_k}$ as

$$\mathbf{h}^{\alpha\alpha}_{\leq j_k} = \begin{pmatrix} \langle\phi_1|\hat{h}|\phi_1\rangle & \langle\phi_1|\hat{h}|\phi_2\rangle & \langle\phi_1|\hat{h}|\phi_3\rangle & \langle\phi_1|\hat{h}|\theta_2\rangle & \langle\phi_1|\hat{h}|\theta_3\rangle & \langle\phi_1|\hat{h}|\phi_6\rangle \\ \langle\phi_2|\hat{h}|\phi_1\rangle & \langle\phi_2|\hat{h}|\phi_2\rangle & \langle\phi_2|\hat{h}|\phi_3\rangle & \langle\phi_2|\hat{h}|\theta_2\rangle & \langle\phi_2|\hat{h}|\theta_3\rangle & \langle\phi_2|\hat{h}|\phi_6\rangle \\ \langle\phi_3|\hat{h}|\phi_1\rangle & \langle\phi_3|\hat{h}|\phi_2\rangle & \langle\phi_3|\hat{h}|\phi_3\rangle & \langle\phi_3|\hat{h}|\theta_2\rangle & \langle\phi_3|\hat{h}|\theta_3\rangle & \langle\phi_3|\hat{h}|\phi_6\rangle \\ \langle\phi_4|\hat{h}|\phi_1\rangle & \langle\phi_4|\hat{h}|\phi_2\rangle & \langle\phi_4|\hat{h}|\phi_3\rangle & \langle\phi_4|\hat{h}|\theta_2\rangle & \langle\phi_4|\hat{h}|\theta_3\rangle & \langle\phi_4|\hat{h}|\phi_6\rangle \\ \langle\phi_5|\hat{h}|\phi_1\rangle & \langle\phi_5|\hat{h}|\phi_2\rangle & \langle\phi_5|\hat{h}|\phi_3\rangle & \langle\phi_5|\hat{h}|\theta_2\rangle & \langle\phi_5|\hat{h}|\theta_3\rangle & \langle\phi_5|\hat{h}|\phi_6\rangle \\ \langle\phi_6|\hat{h}|\phi_1\rangle & \langle\phi_6|\hat{h}|\phi_2\rangle & \langle\phi_6|\hat{h}|\phi_3\rangle & \langle\phi_6|\hat{h}|\theta_2\rangle & \langle\phi_6|\hat{h}|\theta_3\rangle & \langle\phi_6|\hat{h}|\phi_6\rangle \end{pmatrix},$$

$$\mathbf{II}^{\alpha\alpha}_{\leq j_k} = \begin{pmatrix} 1 & 0 & 0 & \langle\phi_1|\theta_2\rangle & \langle\phi_1|\theta_3\rangle & 0 \\ 0 & 1 & 0 & \langle\phi_2|\theta_2\rangle & \langle\phi_2|\theta_3\rangle & 0 \\ 0 & 0 & 1 & \langle\phi_3|\theta_2\rangle & \langle\phi_3|\theta_3\rangle & 0 \\ 0 & 0 & 0 & \langle\phi_4|\theta_2\rangle & \langle\phi_4|\theta_3\rangle & 0 \\ 0 & 0 & 0 & \langle\phi_5|\theta_2\rangle & \langle\phi_5|\theta_3\rangle & 0 \\ 0 & 0 & 0 & \langle\phi_6|\theta_2\rangle & \langle\phi_6|\theta_3\rangle & 1 \end{pmatrix}. \tag{27}$$

Of course, for $\mathbf{h}^{\beta\beta}_{\leq i_k}$ and $\mathbf{II}^{\beta\beta}_{\leq i_k}$ we have similar matrices by using $\langle\beta|\beta\rangle = 1$ and $\langle\theta_i|\theta_j\rangle = \delta_{ij}$.

On the other hand, partially by using a rule of thumb, Pons Viver [22] proposed the following compact formula instead of Eq. (25):

$$\mathcal{F}(z) = \mathrm{tr}\left[\mathbf{h}(z)\cdot\mathrm{adj}(\mathbf{G}(z))\right], \tag{28}$$

where

$$\mathbf{h}(z) = \begin{pmatrix} \mathbf{h}^{\alpha\alpha} & \mathbf{h}^{\alpha\beta} \\ \mathbf{h}^{\beta\alpha} & \mathbf{h}^{\beta\beta}(z) \end{pmatrix}, \tag{29}$$

$$\mathbf{h}^{\alpha\alpha}_{ij} = \langle\phi_i|\hat{h}|\phi_j\rangle, \quad \mathbf{h}^{\alpha\beta}_{ij} = \langle\phi_i|\hat{h}|\theta_j\rangle, \quad \mathbf{h}^{\beta\alpha}_{ij} = \langle\theta_i|\hat{h}|\phi_j\rangle, \quad \mathbf{h}^{\beta\beta}_{ij}(z) = z\cdot\langle\theta_i|\hat{h}|\theta_j\rangle. \tag{30}$$

In the present study, we show that Eq. (28) is equivalent to Eq. (25), as follows. First, we have

$$\begin{aligned}\mathcal{F}(z) &= \mathrm{tr}\left[\mathbf{h}(z)\cdot|\mathbf{G}(z)|\mathbf{G}(z)^{-1}\right] \\ &= \left|z\cdot\mathbf{I}_{v\times v} - \mathbf{C}^{\dagger}\mathbf{C}\right|\cdot\mathrm{tr}\left[\mathbf{h}(z)\cdot\mathbf{G}(z)^{-1}\right] \\ &= \left(\sum_{k=0}^{v} d_k z^{v-k}\right)\cdot\mathrm{tr}\left[\mathbf{h}(z)\cdot\mathbf{G}(z)^{-1}\right] \\ &= \sum_{k=0}^{v} z^{v-k}\left\{d_k\cdot\mathrm{tr}\left[\mathbf{h}(z)\cdot\mathbf{G}(z)^{-1}\right]\right\},\end{aligned} \tag{31}$$

where we have used Eq. (14), Eq. (16), and the following linear algebra formula,

$$\mathrm{adj}(\mathbf{A}) = |\mathbf{A}|\cdot\mathbf{A}^{-1}. \tag{32}$$

Substituting Eq. (18) into Eq. (17), we obtain

$$d_k = (-1)^k \sum_{\substack{1\leq i_1<\cdots<i_k\leq \mu \\ 1\leq j_1<\cdots<j_k\leq \nu}} \begin{vmatrix} \langle \phi_{i_1}|\theta_{j_1}\rangle & \cdots & \langle \phi_{i_1}|\theta_{j_k}\rangle \\ \vdots & \ddots & \vdots \\ \langle \phi_{i_k}|\theta_{j_1}\rangle & \cdots & \langle \phi_{i_k}|\theta_{j_k}\rangle \end{vmatrix} \cdot \begin{vmatrix} \langle \theta_{j_1}|\phi_{i_1}\rangle & \cdots & \langle \theta_{j_1}|\phi_{i_k}\rangle \\ \vdots & \ddots & \vdots \\ \langle \theta_{j_k}|\phi_{i_1}\rangle & \cdots & \langle \theta_{j_k}|\phi_{i_k}\rangle \end{vmatrix}$$

$$= (-1)^k \sum_{\substack{1\leq i_1<\cdots<i_k\leq \mu \\ 1\leq j_1<\cdots<j_k\leq \nu}} \left|\mathbf{S}^{i_k j_k}\right|. \tag{33}$$

Then, substituting Eq. (33) into Eq. (31) gives

$$\mathcal{F}(z) = \sum_{k=0}^{\nu} (-1)^k z^{\nu-k} \sum_{\substack{1\leq i_1<\cdots<i_k\leq \mu \\ 1\leq j_1<\cdots<j_k\leq \nu}} \left|\mathbf{S}^{i_k j_k}\right| \cdot \text{tr}\left[\mathbf{h}(z)\cdot \mathbf{G}(z)^{-1}\right]$$

$$= \sum_{k=0}^{\nu} (-1)^k z^{\nu-k} \sum_{\substack{1\leq i_1<\cdots<i_k\leq \mu \\ 1\leq j_1<\cdots<j_k\leq \nu}} \text{tr}\left[\mathbf{h}(z)\cdot \mathbf{G}(z)^{-1} \cdot \left|\mathbf{S}^{i_k j_k}\right|\right]$$

$$= \sum_{k=0}^{\nu} (-1)^k z^{\nu-k} \sum_{\substack{1\leq i_1<\cdots<i_k\leq \mu \\ 1\leq j_1<\cdots<j_k\leq \nu}} \text{tr}\left[\mathbf{h}^{\text{UHF}} \cdot \mathbf{G}(z) \cdot \mathbf{G}(z)^{-1} \cdot \mathbf{S}^{i_k j_k} \cdot \text{adj}\left(\mathbf{S}^{i_k j_k}\right)\right]$$

$$= \sum_{k=0}^{\nu} (-1)^k z^{\nu-k} \sum_{\substack{1\leq i_1<\cdots<i_k\leq \mu \\ 1\leq j_1<\cdots<j_k\leq \nu}} \text{tr}\left[\mathbf{h}^{\text{UHF}} \cdot \mathbf{S}^{i_k j_k} \cdot \text{adj}\left(\mathbf{S}^{i_k j_k}\right)\right], \tag{34}$$

where we have introduced the following relation,

$$\mathbf{h}(z) = \mathbf{h}^{\text{UHF}} \cdot \mathbf{G}(z) \tag{35}$$

with

$$\mathbf{h}^{\text{UHF}} := \begin{pmatrix} \mathbf{h}^{\alpha\alpha} & \mathbf{0} \\ \mathbf{0} & \mathbf{h}^{\beta\beta} \end{pmatrix} = \begin{pmatrix} \langle \phi_i|\hat{h}|\phi_j\rangle & \mathbf{0} \\ \mathbf{0} & \langle \theta_i|\hat{h}|\theta_j\rangle \end{pmatrix}. \tag{36}$$

This relation, Eq. (35), holds when the following relations are satisfied,

$$\langle \phi_a|\theta_i\rangle = 0 \text{ and } \langle \theta_a|\phi_i\rangle = 0, \tag{37}$$

for $\forall a \in \{\text{virtual orbitals}\}$ and $\forall i \in \{\text{occupied orbitals}\}$. However, generally the condition of Eq. (37) may not be satisfied in the canonical UHF method. Thus, for example, by using the so-called extended pairing theorem [38,39], we must obtain the corresponding orbitals [40–42] satisfying Eq. (37). According to Mayer's paper [39], for closed-shell molecules the condition of Eq. (37) can be satisfied, whereas for open-shell molecules we can have $\langle \phi_a|\theta_i\rangle = 0$ (for $\forall a \in \{\text{virtual orbitals}\}$ and $\forall i \in \{\text{occupied orbitals}\}$), $\langle \theta_a|\phi_i\rangle = 0$ (for $\forall a \in \{\text{virtual orbitals}\}$ and $i = 1, 2, \cdots, \nu$), and $\langle \theta_a|\phi_i\rangle = \delta_{ai}$ (for $i, a = \nu+1, \nu+2, \cdots, \mu$) if we adopt the CO model used in the ROHF method (please see Section 7). Moreover, we can define $\langle \theta_a|\phi_i\rangle = 0$ (for $i, a = \nu+1, \nu+2, \cdots, \mu$), because $\langle \theta_a|\phi_i\rangle = \delta_{ai}$ (for $i, a = \nu+1, \nu+2, \cdots, \mu$) are definitions [39]. Therefore, by using both CO model and extended pairing theorem, we can obtain the corresponding orbitals satisfying Eq. (37) for open-shell molecules. Remarkably, by using both UHF orbitals and extended pairing theorem, we cannot obtain the corresponding orbitals satisfying Eq. (37) for open-shell molecules, because the UHF case has non integer inter-relation overlaps between corresponding orbitals (e.g., $\langle \phi_a | \theta_i \rangle \neq 0$) [39]. That is, the condition of Eq. (37) requires the use of ROHF-like orbitals [43] for the orbitals $\{\phi_p\}$ and $\{\theta_p\}$.

Additionally, the condition of Eq. (37) can convert the orbital completeness relations

$$\sum_{p=1}^{occ+vir} |\phi_p\rangle\langle\phi_p| = 1 \quad \text{and} \quad \sum_{p=1}^{occ+vir} |\theta_p\rangle\langle\theta_p| = 1 \tag{38}$$

to

$$\sum_{j=1}^{occ} |\phi_j\rangle\langle\phi_j|\theta_i\rangle = |\theta_i\rangle \quad \text{and} \quad \sum_{j=1}^{occ} |\theta_j\rangle\langle\theta_j|\phi_i\rangle = |\phi_i\rangle, \tag{39}$$

respectively. Then, by using Eq. (39) we can readily prove Eq. (35) and the following relation:

$$\mathbf{h}^{UHF} \cdot \mathbf{S}^{i_k j_k} = \begin{pmatrix} \mathbf{h}^{\alpha\alpha} \cdot \mathbf{\Pi}^{\alpha\alpha}_{\leq j_k} & 0 \\ 0 & \mathbf{h}^{\beta\beta} \cdot \mathbf{\Pi}^{\beta\beta}_{\leq i_k} \end{pmatrix} = \begin{pmatrix} \mathbf{h}^{\alpha\alpha}_{\leq j_k} & 0 \\ 0 & \mathbf{h}^{\beta\beta}_{\leq i_k} \end{pmatrix} = \mathbf{h}^{i_k j_k}. \tag{40}$$

Therefore, substituting Eq. (40) into Eq. (34) and comparing Eq. (34) with Eq. (25), we obtain

$$\mathcal{F}(z) = \langle \Psi | \sum_{i=1}^{n} \hat{h}(\mathbf{r}_i) | \hat{\mathcal{P}}_S \Psi \rangle = \text{tr}\left[ \mathbf{h}(z) \cdot \text{adj}(\mathbf{G}(z)) \right], \tag{41}$$

where we have used the relations $z^{\nu-k} = C_k(S, M, N)$. We note that we cannot calculate $\mathcal{F}(z)$ of Eq. (28) by using UHF orbitals because of the condition of Eq. (37), while we can calculate $\mathcal{G}(z)$ of Eq. (20) by using UHF orbitals.

## 4. Non-orthogonal matrix elements of two-electron operators

In this section, we consider the following expectation value $\mathcal{V}(C_k)$ of two-electron operators:

$$\mathcal{V}(C_k) = \langle \Psi | \sum_{i<j}^{n} \frac{1}{r_{ij}} | \hat{\mathcal{P}}_S \Psi \rangle = \langle \Psi | \frac{1}{2} \sum_{i}^{n} \sum_{j \neq i}^{n} \frac{1}{r_{ij}} | \hat{\mathcal{P}}_S \Psi \rangle, \tag{42}$$

where $1/r_{ij}$ is the Coulomb interaction between electrons $i$ and $j$. Then, substituting Eqs. (3) and (9) into Eq. (42) gives

$$\mathcal{V}(C_k) = \sum_{k=0}^{\nu} (-1)^k C_k \sum_{\substack{1 \leq i_1 < i_2 < \cdots < i_k \leq \mu \\ 1 \leq j_1 < j_2 < \cdots < j_k \leq \nu}} \langle T_0 | \frac{1}{2} \sum_{i}^{n} \sum_{j \neq i}^{n} \frac{1}{r_{ij}} | T_k^{i_k j_k} \rangle. \tag{43}$$

Generally, the two-electron operator matrix elements between non-orthogonal Slater determinants can be written as [16,36,37]

$$\langle \Phi_F | \sum_{i=1}^{n-1} \sum_{j=i+1}^{n} \frac{1}{r_{ij}} | \Phi_G \rangle = \sum_{i=1}^{n} \sum_{j=i+1}^{n} \sum_{k=1}^{n} \sum_{l=k+1}^{n} V_{ij;kl} \cdot (-1)^{i+j+k+l} \left| \mathbf{S}^{(ij,kl)} \right|$$

$$= \sum_{i=1}^{n} \sum_{j=i+1}^{n} \sum_{k=1}^{n} \sum_{l=k+1}^{n} V_{ij;kl} \cdot \mathrm{adj}^{(2)}(\mathbf{S})_{kl;ij}$$

$$= \mathrm{tr}\left[ \mathbf{V} \cdot \mathrm{adj}^{(2)}(\mathbf{S}) \right], \tag{44}$$

or

$$\langle \Phi_F | \frac{1}{2} \sum_{i=1}^{n} \sum_{j \neq i}^{n} \frac{1}{r_{ij}} | \Phi_G \rangle = \frac{1}{4} \sum_{i=1}^{n} \sum_{j=1}^{n} \sum_{k=1}^{n} \sum_{l=1}^{n} V_{ij;kl} \cdot (-1)^{i+j+k+l} \left| \mathbf{S}^{(ij,kl)} \right|$$

$$= \frac{1}{4} \sum_{i=1}^{n} \sum_{j=1}^{n} \sum_{k=1}^{n} \sum_{l=1}^{n} V_{ij;kl} \cdot \mathrm{adj}^{(2)}(\mathbf{S})_{kl;ij}$$

$$= \frac{1}{4} \mathrm{tr}\left[ \mathbf{V} \cdot \mathrm{adj}^{(2)}(\mathbf{S}) \right], \tag{45}$$

where

$$V_{ij;kl} = \iint f_i^*(\mathbf{r}_1) f_j^*(\mathbf{r}_2) \frac{1}{r_{12}} \left[ g_k(\mathbf{r}_1) g_l(\mathbf{r}_2) - g_l(\mathbf{r}_1) g_k(\mathbf{r}_2) \right] d\mathbf{r}_1 d\mathbf{r}_2$$

$$= \langle ij \| kl \rangle = \langle ij | kl \rangle - \langle ij | lk \rangle, \tag{46}$$

$\left| \mathbf{S}^{(ij,kl)} \right|$ is the determinant obtained from $|\mathbf{S}|$ by deleting the $i$th and $j$th rows ($i<j$) and $k$th and $l$th columns ($k<l$). Thus, by applying Eq. (45) to Eq. (43), we obtain

$$\mathcal{V}(C_k) = \frac{1}{4} \sum_{k=0}^{\nu} (-1)^k C_k \sum_{\substack{1 \leq i_1 < i_2 < \cdots < i_k \leq \mu \\ 1 \leq j_1 < j_2 < \cdots < j_k \leq \nu}} \sum_{i=1}^{n} \sum_{j=1}^{n} \sum_{k=1}^{n} \sum_{l=1}^{n} \langle ij \| \breve{k}\breve{l} \rangle^{i_k j_k} \cdot (-1)^{i+j+k+l} \left| \mathbf{S}^{i_k j_k (ij,kl)} \right| \tag{47}$$

$$= \frac{1}{4} \sum_{k=0}^{\nu} (-1)^k C_k \sum_{\substack{1 \leq i_1 < i_2 < \cdots < i_k \leq \mu \\ 1 \leq j_1 < j_2 < \cdots < j_k \leq \nu}} \mathrm{tr}\left[ \mathbf{V}^{i_k j_k} \cdot \mathrm{adj}^{(2)}\left( \mathbf{S}^{i_k j_k} \right) \right], \tag{48}$$

where

$$\mathbf{V}^{i_k j_k} = \begin{pmatrix} \breve{V}_{ij\|kl}^{\alpha\alpha;\alpha\alpha} & \breve{V}_{ij\|kl}^{\alpha\alpha;\alpha\beta} & \breve{V}_{ij\|kl}^{\alpha\alpha;\beta\alpha} & \breve{V}_{ij\|kl}^{\alpha\alpha;\beta\beta} \\ \breve{V}_{ij\|kl}^{\alpha\beta;\alpha\alpha} & \breve{V}_{ij\|kl}^{\alpha\beta;\alpha\beta} & \breve{V}_{ij\|kl}^{\alpha\beta;\beta\alpha} & \breve{V}_{ij\|kl}^{\alpha\beta;\beta\beta} \\ \breve{V}_{ij\|kl}^{\beta\alpha;\alpha\alpha} & \breve{V}_{ij\|kl}^{\beta\alpha;\alpha\beta} & \breve{V}_{ij\|kl}^{\beta\alpha;\beta\alpha} & \breve{V}_{ij\|kl}^{\beta\alpha;\beta\beta} \\ \breve{V}_{ij\|kl}^{\beta\beta;\alpha\alpha} & \breve{V}_{ij\|kl}^{\beta\beta;\alpha\beta} & \breve{V}_{ij\|kl}^{\beta\beta;\beta\alpha} & \breve{V}_{ij\|kl}^{\beta\beta;\beta\beta} \end{pmatrix} = \begin{pmatrix} \breve{V}_{ij\|kl}^{\alpha\alpha;\alpha\alpha} & 0 & 0 & 0 \\ 0 & \breve{V}_{ij\|kl}^{\alpha\beta;\alpha\beta} & \breve{V}_{-ij\|lk}^{\alpha\beta;\alpha\beta} & 0 \\ 0 & \breve{V}_{-ij\|lk}^{\beta\alpha;\beta\alpha} & \breve{V}_{ij\|kl}^{\beta\alpha;\beta\alpha} & 0 \\ 0 & 0 & 0 & \breve{V}_{ij\|kl}^{\beta\beta;\beta\beta} \end{pmatrix}$$

$$= \begin{pmatrix} \langle f_i^\alpha f_j^\alpha \| \breve{g}_k^\alpha \breve{g}_l^\alpha \rangle & 0 & 0 & 0 \\ 0 & \langle f_i^\alpha f_j^\beta | \breve{g}_k^\alpha \breve{g}_l^\beta \rangle & -\langle f_i^\alpha f_j^\beta | \breve{g}_l^\alpha \breve{g}_k^\beta \rangle & 0 \\ 0 & -\langle f_i^\beta f_j^\alpha | \breve{g}_l^\beta \breve{g}_k^\alpha \rangle & \langle f_i^\beta f_j^\alpha | \breve{g}_k^\beta \breve{g}_l^\alpha \rangle & 0 \\ 0 & 0 & 0 & \langle f_i^\beta f_j^\beta \| \breve{g}_k^\beta \breve{g}_l^\beta \rangle \end{pmatrix} \tag{49}$$

$$= \begin{pmatrix} \langle \phi_i \phi_j \| \breve{g}_k' \breve{g}_l' \rangle & 0 & 0 & 0 \\ 0 & \langle \phi_i \theta_j | \breve{g}_k' \breve{g}_l' \rangle & -\langle \phi_i \theta_j | \breve{g}_l' \breve{g}_k' \rangle & 0 \\ 0 & -\langle \theta_i \phi_j | \breve{g}_l' \breve{g}_k' \rangle & \langle \theta_i \phi_j | \breve{g}_k' \breve{g}_l' \rangle & 0 \\ 0 & 0 & 0 & \langle \theta_i \theta_j \| \breve{g}_k' \breve{g}_l' \rangle \end{pmatrix}. \tag{50}$$

The $|\breve{k}\breve{l}\rangle$ part of $\langle ij \| \breve{k}\breve{l} \rangle^{i_k j_k}$ in Eq. (47) indicates that the $|\breve{k}\breve{l}\rangle$ part is composed of spin-orbitals in the Slater determinant $|T_k^{i_k j_k}\rangle$, and the $|\breve{g}_k^\alpha \breve{g}_l^\beta\rangle$ part in Eq. (49) also has the same meaning as $|\breve{k}\breve{l}\rangle$. The superscript $\alpha$ of the spin-orbitals $f_i^\alpha$ and $\breve{g}_k^\alpha$ in Eq. (49) emphasizes that the spin-orbitals $f_i$ and $\breve{g}_k$ possess the spin function $|\alpha\rangle$, while $\breve{g}_k'$ in Eq. (50) denotes the spatial part of the spin-orbital $\breve{g}_k^\alpha$. The spatial orbital $\breve{g}_k'$ is either $\phi_k$ or $\theta_k$. The zero matrices $\mathbf{0}$ in the matrix $\mathbf{V}^{i_k j_k}$ are attributed to $\langle\alpha|\beta\rangle = 0$ or $\langle\beta|\alpha\rangle = 0$.

On the other hand, partially by using a rule of thumb, Pons Viver [22] proposed the following compact formula instead of Eq. (48):

$$\mathcal{V}(z) = \frac{1}{4}\mathrm{tr}\left[\mathbf{V}(z)\cdot\mathrm{adj}^{(2)}\left(\mathbf{G}(z)\right)\right], \tag{51}$$

where

$$\mathbf{V}(z) = \begin{pmatrix} \langle\phi_i\phi_j\|\phi_k\phi_l\rangle & \langle\phi_i\phi_j\|\phi_k\theta_l\rangle & \langle\phi_i\phi_j\|\theta_k\phi_l\rangle & \langle\phi_i\phi_j\|\theta_k\theta_l\rangle \\ \langle\phi_i\theta_j\|\phi_k\phi_l\rangle & z\langle\phi_i\phi_j|\phi_k\theta_l\rangle - \langle\phi_i\theta_j|\theta_l\phi_k\rangle & \langle\phi_i\phi_j|\theta_k\phi_l\rangle - z\langle\phi_i\theta_j|\phi_l\theta_k\rangle & z\langle\phi_i\theta_j\|\theta_k\theta_l\rangle \\ \langle\theta_i\phi_j\|\phi_k\phi_l\rangle & \langle\theta_i\phi_j|\phi_k\theta_l\rangle - z\langle\theta_i\phi_j|\theta_l\phi_k\rangle & z\langle\theta_i\phi_j|\theta_k\phi_l\rangle - \langle\phi_i\phi_j|\phi_l\theta_k\rangle & z\langle\theta_i\phi_j\|\theta_k\theta_l\rangle \\ \langle\theta_i\theta_j\|\phi_k\phi_l\rangle & z\langle\theta_i\theta_j\|\phi_k\theta_l\rangle & z\langle\theta_i\theta_j\|\theta_k\phi_l\rangle & z^2\langle\theta_i\theta_j\|\theta_k\theta_l\rangle \end{pmatrix}. \tag{52}$$

In the present study, we show that Eq. (51) is equivalent to Eq. (48), as follows. First, we have

$$\begin{aligned}\mathcal{V}(z) &= \frac{1}{4}\mathrm{tr}\left[\mathbf{V}(z)\cdot|\mathbf{G}(z)|\cdot\left(\mathbf{G}^{(2)}(z)\right)^{-1}\right] \\ &= \frac{1}{4}\left|z\cdot\mathbf{I}_\nu - \mathbf{C}^\dagger\mathbf{C}\right|\cdot\mathrm{tr}\left[\mathbf{V}(z)\cdot\left(\mathbf{G}^{(2)}(z)\right)^{-1}\right] \\ &= \frac{1}{4}\left(\sum_{k=0}^\nu d_k z^{\nu-k}\right)\cdot\mathrm{tr}\left[\mathbf{V}(z)\cdot\left(\mathbf{G}^{(2)}(z)\right)^{-1}\right] \\ &= \frac{1}{4}\sum_{k=0}^\nu z^{\nu-k}\left\{d_k\cdot\mathrm{tr}\left[\mathbf{V}(z)\cdot\left(\mathbf{G}^{(2)}(z)\right)^{-1}\right]\right\},\end{aligned} \tag{53}$$

where we have used Eq. (14), Eq. (16), and the following linear algebra formula,

$$\mathrm{adj}^{(2)}(\mathbf{A}) = |\mathbf{A}|\cdot\left(\mathbf{A}^{(2)}\right)^{-1}. \tag{54}$$

Here, $\mathrm{adj}^{(2)}(\mathbf{A})$ and $\mathbf{A}^{(2)}$ are the second-order adjugate matrix and the second-order compound matrix of a matrix $\mathbf{A} = (a_{ij})$, respectively, and the matrix elements of $\mathbf{A}^{(2)}$ are defined by $A^{(2)}_{ij;kl} = a_{ik}a_{jl} - a_{il}a_{jk}$ ($i < j$ and $k < l$).

Then, substituting Eq. (33) into Eq. (53) gives

$$\mathcal{V}(z) = \frac{1}{4}\sum_{k=0}^{\nu}(-1)^k z^{\nu-k} \sum_{\substack{1\leq i_1<i_2<\cdots<i_k\leq \mu \\ 1\leq j_1<j_2<\cdots<j_k\leq \nu}} \left|\mathbf{S}^{i_k j_k}\right| \mathrm{tr}\left[\mathbf{V}(z)\cdot\left(\mathbf{G}^{(2)}(z)\right)^{-1}\right]$$

$$= \frac{1}{4}\sum_{k=0}^{\nu}(-1)^k z^{\nu-k} \sum_{\substack{1\leq i_1<i_2<\cdots<i_k\leq \mu \\ 1\leq j_1<j_2<\cdots<j_k\leq \nu}} \mathrm{tr}\left[\mathbf{V}(z)\cdot\left(\mathbf{G}^{(2)}(z)\right)^{-1}\cdot\left|\mathbf{S}^{i_k j_k}\right|\right]$$

$$= \frac{1}{4}\sum_{k=0}^{\nu}(-1)^k z^{\nu-k} \sum_{\substack{1\leq i_1<i_2<\cdots<i_k\leq \mu \\ 1\leq j_1<j_2<\cdots<j_k\leq \nu}} \mathrm{tr}\left[\mathbf{V}^{\mathrm{UHF}}\cdot\mathbf{G}^{(2)}(z)\cdot\left(\mathbf{G}^{(2)}(z)\right)^{-1}\cdot\mathbf{S}^{i_k j_k(2)}\cdot\mathrm{adj}^{(2)}\left(\mathbf{S}^{i_k j_k}\right)\right]$$

$$= \frac{1}{4}\sum_{k=0}^{\nu}(-1)^k z^{\nu-k} \sum_{\substack{1\leq i_1<i_2<\cdots<i_k\leq \mu \\ 1\leq j_1<j_2<\cdots<j_k\leq \nu}} \mathrm{tr}\left[\mathbf{V}^{\mathrm{UHF}}\cdot\mathbf{S}^{i_k j_k(2)}\cdot\mathrm{adj}^{(2)}\left(\mathbf{S}^{i_k j_k}\right)\right], \tag{55}$$

where we have introduced the following relation,

$$\mathbf{V}(z) = \mathbf{V}^{\mathrm{UHF}}\cdot\mathbf{G}^{(2)}(z) \tag{56}$$

with

$$\mathbf{V}^{\mathrm{UHF}} := \begin{pmatrix} V^{\alpha\alpha;\alpha\alpha}_{ij\|kl} & V^{\alpha\alpha;\alpha\beta}_{ij\|kl} & V^{\alpha\alpha;\beta\alpha}_{ij\|kl} & V^{\alpha\alpha;\beta\beta}_{ij\|kl} \\ V^{\alpha\beta;\alpha\alpha}_{ij\|kl} & V^{\alpha\beta;\alpha\beta}_{ij\|kl} & V^{\alpha\beta;\beta\alpha}_{ij\|kl} & V^{\alpha\beta;\beta\beta}_{ij\|kl} \\ V^{\beta\alpha;\alpha\alpha}_{ij\|kl} & V^{\beta\alpha;\alpha\beta}_{ij\|kl} & V^{\beta\alpha;\beta\alpha}_{ij\|kl} & V^{\beta\alpha;\beta\beta}_{ij\|kl} \\ V^{\beta\beta;\alpha\alpha}_{ij\|kl} & V^{\beta\beta;\alpha\beta}_{ij\|kl} & V^{\beta\beta;\beta\alpha}_{ij\|kl} & V^{\beta\beta;\beta\beta}_{ij\|kl} \end{pmatrix} = \begin{pmatrix} V^{\alpha\alpha;\alpha\alpha}_{ij\|kl} & 0 & 0 & 0 \\ 0 & V^{\alpha\beta;\alpha\beta}_{ij\|kl} & V^{\alpha\beta;\alpha\beta}_{-ij\|lk} & 0 \\ 0 & V^{\beta\alpha;\beta\alpha}_{-ij\|lk} & V^{\beta\alpha;\beta\alpha}_{ij\|kl} & 0 \\ 0 & 0 & 0 & V^{\beta\beta;\beta\beta}_{ij\|kl} \end{pmatrix}$$

$$= \begin{pmatrix} \langle\phi_i\phi_j\|\phi_k\phi_l\rangle & 0 & 0 & 0 \\ 0 & \langle\phi_i\theta_j|\phi_k\theta_l\rangle & -\langle\phi_i\theta_j|\phi_l\theta_k\rangle & 0 \\ 0 & -\langle\theta_i\phi_j|\theta_l\phi_k\rangle & \langle\theta_i\phi_j|\theta_k\phi_l\rangle & 0 \\ 0 & 0 & 0 & \langle\theta_i\theta_j\|\theta_k\theta_l\rangle \end{pmatrix}. \tag{57}$$

Next, by using Eq. (39) we prove the relation, $\mathbf{V}(z) = \mathbf{V}^{\mathrm{UHF}}\cdot\mathbf{G}^{(2)}(z)$, as follows. First, we consider the following arbitrary matrix element of $\mathbf{V}^{\mathrm{UHF}}\cdot\mathbf{G}^{(2)}(z)$:

$$\left(\mathbf{V}^{\mathrm{UHF}}\cdot\mathbf{G}^{(2)}(z)\right)_{IJ;ij} = \sum_{k=1}^{n}\sum_{l=k+1}^{n} V^{\mathrm{UHF}}_{IJ;kl}\cdot G^{(2)}_{kl;ij}(z)$$

$$= \frac{1}{2}\sum_{k=1}^{n}\sum_{l=1}^{n} V^{\mathrm{UHF}}_{IJ;kl}\cdot G^{(2)}_{kl;ij}(z)$$

$$= \frac{1}{2}\sum_{k=1}^{n}\sum_{l=1}^{n}\langle IJ\|kl\rangle\left(g_{ki}g_{lj}-g_{kj}g_{li}\right)$$

$$= \frac{1}{2}\sum_{k=1}^{n}\sum_{l=1}^{n}\left(\langle IJ\|kl\rangle g_{ki}g_{lj}+\langle JI\|kl\rangle g_{kj}g_{li}\right)$$

$$= \frac{1}{2}\sum_{k=1}^{n}\sum_{l=1}^{n}\left(\langle IJ\|kl\rangle g_{ki}g_{lj}+\langle IJ\|lk\rangle g_{li}g_{kj}\right)$$

$$= \sum_{k=1}^{n}\sum_{l=1}^{n}\langle IJ\|kl\rangle g_{ki}g_{lj}, \tag{58}$$

where $g_{ij}$ are the matrix elements of $\mathbf{G}(z)$: $\mathbf{G}(z) = (g_{ij})$. Then, substituting

$$\sum_{k=1}^{n}\sum_{l=1}^{n} = \sum_{k=1}^{\mu}\sum_{l=1}^{\mu} + \sum_{k=1}^{\mu}\sum_{l=\mu+1}^{n} + \sum_{k=\mu+1}^{n}\sum_{l=1}^{\mu} + \sum_{k=\mu+1}^{n}\sum_{l=\mu+1}^{n}, \tag{59}$$

into Eq. (58) gives

$$\left(\mathbf{V}^{\text{UHF}} \cdot \mathbf{G}^{(2)}(z)\right)_{IJ;ij} = \sum_{k=1}^{\mu}\sum_{l=1}^{\mu}\langle IJ \| kl\rangle g_{ki}g_{lj} + \sum_{k=1}^{\mu}\sum_{l=\mu+1}^{n}\langle IJ \| kl\rangle g_{ki}g_{lj}$$
$$+ \sum_{k=\mu+1}^{n}\sum_{l=1}^{\mu}\langle IJ \| kl\rangle g_{ki}g_{lj} + \sum_{k=\mu+1}^{n}\sum_{l=\mu+1}^{n}\langle IJ \| kl\rangle g_{ki}g_{lj}. \quad (60)$$

Since the four independent subscripts $I$, $J$, $i$, and $j$ of $\left(\mathbf{V}^{\text{UHF}} \cdot \mathbf{G}^{(2)}(z)\right)_{IJ;ij}$ run all the occupied spin-orbitals with the spin function $|\alpha\rangle$ or $|\beta\rangle$, it is convenient to introduce the following matrix $\mathbf{Q}$ composed of block matrices:

$$\mathbf{Q} := \mathbf{V}^{\text{UHF}} \cdot \mathbf{G}^{(2)}(z) = \begin{pmatrix} \mathbf{Q}_{IJ;ij}^{\alpha\alpha;\alpha\alpha} & \mathbf{Q}_{IJ;ij}^{\alpha\alpha;\alpha\beta} & \mathbf{Q}_{IJ;ij}^{\alpha\alpha;\beta\alpha} & \mathbf{Q}_{IJ;ij}^{\alpha\alpha;\beta\beta} \\ \mathbf{Q}_{IJ;ij}^{\alpha\beta;\alpha\alpha} & \mathbf{Q}_{IJ;ij}^{\alpha\beta;\alpha\beta} & \mathbf{Q}_{IJ;ij}^{\alpha\beta;\beta\alpha} & \mathbf{Q}_{IJ;ij}^{\alpha\beta;\beta\beta} \\ \mathbf{Q}_{IJ;ij}^{\beta\alpha;\alpha\alpha} & \mathbf{Q}_{IJ;ij}^{\beta\alpha;\alpha\beta} & \mathbf{Q}_{IJ;ij}^{\beta\alpha;\beta\alpha} & \mathbf{Q}_{IJ;ij}^{\beta\alpha;\beta\beta} \\ \mathbf{Q}_{IJ;ij}^{\beta\beta;\alpha\alpha} & \mathbf{Q}_{IJ;ij}^{\beta\beta;\alpha\beta} & \mathbf{Q}_{IJ;ij}^{\beta\beta;\beta\alpha} & \mathbf{Q}_{IJ;ij}^{\beta\beta;\beta\beta} \end{pmatrix}, \quad (61)$$

where, for example, we have

$$\mathbf{Q}_{IJ;ij}^{\beta\alpha;\alpha\beta} = \sum_{k=1}^{\mu}\sum_{l=1}^{\mu}\langle I^{\beta}J^{\alpha} \| k^{\alpha}l^{\alpha}\rangle g_{ki}^{\alpha\alpha} g_{lj}^{\alpha\beta} + \sum_{k=1}^{\mu}\sum_{l=1}^{\nu}\langle I^{\beta}J^{\alpha} \| k^{\alpha}l^{\beta}\rangle g_{ki}^{\alpha\alpha} g_{lj}^{\beta\beta}$$
$$+ \sum_{k=1}^{\nu}\sum_{l=1}^{\mu}\langle I^{\beta}J^{\alpha} \| k^{\beta}l^{\alpha}\rangle g_{ki}^{\beta\alpha} g_{lj}^{\alpha\beta} + \sum_{k=1}^{\nu}\sum_{l=1}^{\nu}\langle I^{\beta}J^{\alpha} \| k^{\beta}l^{\beta}\rangle g_{ki}^{\beta\alpha} g_{lj}^{\beta\beta}, \quad (62)$$

with the definitions,

$$\begin{pmatrix} (g_{ij}^{\alpha\alpha}) & (g_{ij}^{\alpha\beta}) \\ (g_{ij}^{\beta\alpha}) & (g_{ij}^{\beta\beta}) \end{pmatrix} := \begin{pmatrix} \mathbf{I}_{\mu\times\mu} & \mathbf{C} \\ \mathbf{C}^{\dagger} & z\mathbf{I}_{\nu\times\nu} \end{pmatrix} = \mathbf{G}(z). \quad (63)$$

For example, calculating $\mathbf{Q}_{IJ;ij}^{\beta\alpha;\alpha\beta}$ by using Eq. (39), we obtain

$$\mathbf{Q}_{IJ;ij}^{\beta\alpha;\alpha\beta} = \sum_{k=1}^{\mu}\sum_{l=1}^{\nu}\langle I^{\beta}J^{\alpha} \| k^{\alpha}l^{\beta}\rangle g_{ki}^{\alpha\alpha} g_{lj}^{\beta\beta} + \sum_{k=1}^{\nu}\sum_{l=1}^{\mu}\langle I^{\beta}J^{\alpha} \| k^{\beta}l^{\alpha}\rangle g_{ki}^{\beta\alpha} g_{lj}^{\alpha\beta}$$
$$= -\sum_{k=1}^{\mu}\sum_{l=1}^{\nu}\langle IJ | lk\rangle \delta_{ki} \cdot z\delta_{lj} + \sum_{k=1}^{\nu}\sum_{l=1}^{\mu}\langle IJ | kl\rangle\langle k|i\rangle\langle l|j\rangle \quad (64)$$
$$= -z\langle IJ | ji\rangle + \langle IJ | ij\rangle$$
$$= \langle\theta_I\phi_J | \phi_i\theta_j\rangle - z\langle\theta_I\phi_J | \theta_j\phi_i\rangle, \quad (65)$$

where we note that the two-electron integrals $\langle IJ | lk\rangle$ and $\langle IJ | kl\rangle$ in Eq. (64) have been described by the spatial orbitals. Similarly calculating all the remaining elements of $\mathbf{Q}$, we finally obtain $\mathbf{Q} = \mathbf{V}(z)$. Therefore, the relation, $\mathbf{V}(z) = \mathbf{V}^{\text{UHF}} \cdot \mathbf{G}^{(2)}(z)$, holds.

Next, by using Eq. (39) we prove the relation, $\mathbf{V}^{\text{UHF}} \cdot \mathbf{S}^{i_k j_k (2)} = \mathbf{V}^{i_k j_k}$, as follows. First, we consider the following arbitrary matrix element of $\mathbf{V}^{\text{UHF}} \cdot \mathbf{S}^{i_k j_k (2)}$:

$$\left(\mathbf{V}^{\text{UHF}}\cdot\mathbf{S}^{i_kj_k(2)}\right)_{IJ;ij} = \sum_{k=1}^{n}\sum_{l=k+1}^{n} V_{IJ;kl}^{\text{UHF}} \cdot \mathbf{S}_{kl;ij}^{i_kj_k(2)}$$

$$= \frac{1}{2}\sum_{k=1}^{n}\sum_{l=1}^{n} V_{IJ;kl}^{\text{UHF}} \cdot \mathbf{S}_{kl;ij}^{i_kj_k(2)}$$

$$= \frac{1}{2}\sum_{k=1}^{n}\sum_{l=1}^{n} \langle IJ\|kl\rangle \left(s_{ki}s_{lj} - s_{kj}s_{li}\right)$$

$$= \frac{1}{2}\sum_{k=1}^{n}\sum_{l=1}^{n} \left(\langle IJ\|kl\rangle s_{ki}s_{lj} + \langle JI\|kl\rangle s_{kj}s_{li}\right)$$

$$= \frac{1}{2}\sum_{k=1}^{n}\sum_{l=1}^{n} \left(\langle IJ\|kl\rangle s_{ki}s_{lj} + \langle IJ\|lk\rangle s_{li}s_{kj}\right)$$

$$= \sum_{k=1}^{n}\sum_{l=1}^{n} \langle IJ\|kl\rangle s_{ki}s_{lj}, \tag{66}$$

where $s_{ij}$ are the matrix elements of $\mathbf{S}^{i_kj_k}$: $\mathbf{S}^{i_kj_k} = (s_{ij})$. Then, substituting Eq. (59) into Eq. (66) gives

$$\left(\mathbf{V}^{\text{UHF}}\cdot\mathbf{S}^{i_kj_k(2)}\right)_{IJ;ij} = \sum_{k=1}^{\mu}\sum_{l=1}^{\mu}\langle IJ\|kl\rangle s_{ki}s_{lj} + \sum_{k=1}^{\mu}\sum_{l=\mu+1}^{n}\langle IJ\|kl\rangle s_{ki}s_{lj}$$

$$+ \sum_{k=\mu+1}^{n}\sum_{l=1}^{\mu}\langle IJ\|kl\rangle s_{ki}s_{lj} + \sum_{k=\mu+1}^{n}\sum_{l=\mu+1}^{n}\langle IJ\|kl\rangle s_{ki}s_{lj}. \tag{67}$$

Since the four independent subscripts $I$, $J$, $i$, and $j$ of $\left(\mathbf{V}^{\text{UHF}}\cdot\mathbf{S}^{i_kj_k(2)}\right)_{IJ;ij}$ run all the occupied spin-orbitals with the spin function $|\alpha\rangle$ or $|\beta\rangle$, it is convenient to introduce the following matrix $\mathbf{U}$ composed of block matrices:

$$\mathbf{U} := \mathbf{V}^{\text{UHF}}\cdot\mathbf{S}^{i_kj_k(2)} = \begin{pmatrix} \mathbf{U}_{IJ;ij}^{\alpha\alpha;\alpha\alpha} & \mathbf{U}_{IJ;ij}^{\alpha\alpha;\alpha\beta} & \mathbf{U}_{IJ;ij}^{\alpha\alpha;\beta\alpha} & \mathbf{U}_{IJ;ij}^{\alpha\alpha;\beta\beta} \\ \mathbf{U}_{IJ;ij}^{\alpha\beta;\alpha\alpha} & \mathbf{U}_{IJ;ij}^{\alpha\beta;\alpha\beta} & \mathbf{U}_{IJ;ij}^{\alpha\beta;\beta\alpha} & \mathbf{U}_{IJ;ij}^{\alpha\beta;\beta\beta} \\ \mathbf{U}_{IJ;ij}^{\beta\alpha;\alpha\alpha} & \mathbf{U}_{IJ;ij}^{\beta\alpha;\alpha\beta} & \mathbf{U}_{IJ;ij}^{\beta\alpha;\beta\alpha} & \mathbf{U}_{IJ;ij}^{\beta\alpha;\beta\beta} \\ \mathbf{U}_{IJ;ij}^{\beta\beta;\alpha\alpha} & \mathbf{U}_{IJ;ij}^{\beta\beta;\alpha\beta} & \mathbf{U}_{IJ;ij}^{\beta\beta;\beta\alpha} & \mathbf{U}_{IJ;ij}^{\beta\beta;\beta\beta} \end{pmatrix}, \tag{68}$$

where, for example, we have

$$\mathbf{U}_{IJ;ij}^{\beta\alpha;\alpha\beta} = \sum_{k=1}^{\mu}\sum_{l=1}^{\mu}\langle I^\beta J^\alpha \| k^\alpha l^\alpha\rangle s_{ki}^{\alpha\alpha} s_{lj}^{\alpha\beta} + \sum_{k=1}^{\mu}\sum_{l=1}^{\nu}\langle I^\beta J^\alpha \| k^\alpha l^\beta\rangle s_{ki}^{\alpha\alpha} s_{lj}^{\beta\beta}$$

$$+ \sum_{k=1}^{\nu}\sum_{l=1}^{\mu}\langle I^\beta J^\alpha \| k^\beta l^\alpha\rangle s_{ki}^{\beta\alpha} s_{lj}^{\alpha\beta} + \sum_{k=1}^{\nu}\sum_{l=1}^{\nu}\langle I^\beta J^\alpha \| k^\beta l^\beta\rangle s_{ki}^{\beta\alpha} s_{lj}^{\beta\beta}. \tag{69}$$

with the definitions,

$$\begin{pmatrix} (s_{ij}^{\alpha\alpha}) & (s_{ij}^{\alpha\beta}) \\ (s_{ij}^{\beta\alpha}) & (s_{ij}^{\beta\beta}) \end{pmatrix} := \begin{pmatrix} \mathbf{II}_{\leq j_k}^{\alpha\alpha} & \mathbf{0} \\ \mathbf{0} & \mathbf{II}_{\leq i_k}^{\beta\beta} \end{pmatrix} = \mathbf{S}^{i_kj_k}. \tag{70}$$

For example, calculating $\mathbf{U}_{IJ;ij}^{\beta\alpha;\alpha\beta}$ by using Eq. (39), we have

$$U_{IJ;ij}^{\beta\alpha;\alpha\beta} = \sum_{k=1}^{\mu}\sum_{l=1}^{\nu}\langle I^{\beta}J^{\alpha}\|k^{\alpha}l^{\beta}\rangle s_{ki}^{\alpha\alpha}s_{lj}^{\beta\beta} + \sum_{k=1}^{\nu}\sum_{l=1}^{\mu}\langle I^{\beta}J^{\alpha}\|k^{\beta}l^{\alpha}\rangle s_{ki}^{\beta\alpha}s_{lj}^{\alpha\beta}$$

$$= -\sum_{k=1}^{\mu}\sum_{l=1}^{\nu}\langle IJ|lk\rangle\langle k|\breve{i}\rangle\langle l|\breve{j}\rangle \tag{71}$$

$$= -\langle IJ|\breve{j}\breve{i}\rangle = -\langle \theta_I \phi_J | \breve{g}'_j \breve{g}'_i \rangle, \tag{72}$$

where we note that the two-electron integrals $\langle IJ|lk\rangle$ in Eq. (71) have been described by the spatial orbitals. Similarly calculating all the remaining elements of $\mathbf{U}$, we finally obtain $\mathbf{U} = \mathbf{V}^{i_k j_k}$. Hence, from Eq. (68) the relation,

$$\mathbf{V}^{i_k j_k} = \mathbf{V}^{\text{UHF}} \cdot \mathbf{S}^{i_k j_k (2)}, \tag{73}$$

holds. Substituting Eq. (73) into Eq. (55) and comparing Eq. (55) with Eq. (48), we finally obtain

$$\mathcal{V}(z) = \langle \Psi | \sum_{i<j}^{n} \frac{1}{r_{ij}} | \hat{\mathcal{P}}_S \Psi \rangle = \frac{1}{4}\text{tr}\left[\mathbf{V}(z) \cdot \text{adj}^{(2)}(\mathbf{G}(z))\right], \tag{74}$$

where we have used the relations $z^{\nu-k} = C_k(S,M,N)$. We note that the prefactor 1/4 in Eq. (74) can be removed by the definition of the trace of the matrix product $\mathbf{V}(z) \cdot \text{adj}^{(2)}(\mathbf{G}(z))$ (please see Eqs. (44) and (45)).

Incidentally, we can prove that $\left(\mathbf{G}^{(2)}(z)\right)^{-1}$ exists, as follows. We have $\left(\mathbf{G}^{(2)}(z)\right)^{-1} = \left(\mathbf{G}(z)^{-1}\right)^{(2)}$ [44,45]. Thus, if $\mathbf{G}(z)^{-1}$ exists, $\left(\mathbf{G}^{(2)}(z)\right)^{-1}$ exists. When $|\mathbf{G}(z)| \neq 0$, $\mathbf{G}(z)^{-1}$ exists. Since $|\mathbf{G}(z)| = \langle \Psi|\hat{\mathcal{P}}_S|\Psi\rangle$, we have $|\mathbf{G}(z)| \neq 0$ if $\langle \Psi|\hat{\mathcal{P}}_S|\Psi\rangle \neq 0$. Therefore, if $\langle \Psi|\hat{\mathcal{P}}_S|\Psi\rangle \neq 0$, $\left(\mathbf{G}^{(2)}(z)\right)^{-1}$ exists. From Eq. (1) it is obvious that we have assumed $\langle \Psi|\hat{\mathcal{P}}_S|\Psi\rangle \neq 0$.

## 5. Adjugate matrix of $\tilde{\mathbf{G}}(z)$

As the preparation for simplifying the explicit expressions of $\mathcal{F}(z)$ and $\mathcal{V}(z)$, we derive the adjugate of $\mathbf{G}(z)$ (strictly speaking, $\tilde{\mathbf{G}}(z)$ of Eq. (76)) by adopting the corresponding orbitals. (We note that, strictly speaking, only for closed-shell molecules the following derivations should hold, because we do not adopt the CO model in this section.) By using the unitary matrix $\mathbf{T}$ of the pairing theorem [22], the expressions of $\mathcal{F}(z)$ and $\mathcal{V}(z)$ can be converted to

$$\mathcal{F}(z) + \mathcal{V}(z) = \text{tr}\left[\mathbf{T}^{\text{T}} \cdot \mathbf{h}(z) \cdot \mathbf{T} \cdot \text{adj}\left(\mathbf{T}^{\text{T}} \cdot \mathbf{G}(z) \cdot \mathbf{T}\right)\right] + \frac{1}{4}\text{tr}\left[\left(\mathbf{T}^{(2)}\right)^{\text{T}} \cdot \mathbf{V}(z) \cdot \mathbf{T}^{(2)} \cdot \text{adj}^{(2)}\left(\mathbf{T}^{\text{T}} \cdot \mathbf{G}(z) \cdot \mathbf{T}\right)\right]$$

$$= \text{tr}\left[\tilde{\mathbf{h}}(z) \cdot \text{adj}(\tilde{\mathbf{G}}(z))\right] + \frac{1}{4}\text{tr}\left[\tilde{\mathbf{V}}(z) \cdot \text{adj}^{(2)}(\tilde{\mathbf{G}}(z))\right], \tag{75}$$

where $\tilde{\mathbf{h}}(z) := \mathbf{T}^{\text{T}} \cdot \mathbf{h}(z) \cdot \mathbf{T}$, $\tilde{\mathbf{V}}(z) := \left(\mathbf{T}^{(2)}\right)^{\text{T}} \cdot \mathbf{V}(z) \cdot \mathbf{T}^{(2)}$, and

$$\tilde{\mathbf{G}}(z) := \mathbf{T}^T \cdot \mathbf{G}(z) \cdot \mathbf{T} = \begin{pmatrix} \mathbf{I}_{\mu\times\mu} & \begin{matrix} \gamma_1 & 0 & \cdots & 0 \\ 0 & \gamma_2 & & 0 \\ \vdots & & \ddots & \vdots \\ 0 & \cdots & \cdots & \gamma_\nu \\ 0 & 0 & \cdots & 0 \\ 0 & 0 & \cdots & 0 \end{matrix} \\ \begin{matrix} \gamma_1^* & 0 & \cdots & 0 & 0 & 0 \\ 0 & \gamma_2^* & & \vdots & 0 & 0 \\ \vdots & & \ddots & \vdots & \vdots & \vdots \\ 0 & 0 & \cdots & \gamma_\nu^* & 0 & 0 \end{matrix} & z\mathbf{I}_{\nu\times\nu} \end{pmatrix}, \tag{76}$$

with $\gamma_i = \langle \phi_i | \theta_i \rangle$ and $\gamma_i^* = \langle \theta_i | \phi_i \rangle$ [22]. We note that when calculating $\mathcal{F}(z)$ and $\mathcal{V}(z)$, we calculate $\mathbf{h}(z)$ and $\mathbf{V}(z)$ by using the corresponding orbitals, instead of calculating $\tilde{\mathbf{h}}(z)$ and $\tilde{\mathbf{V}}(z)$ by using the canonical orbitals [12]. Then, the adjugate of $\tilde{\mathbf{G}}(z)$ becomes

$$\begin{aligned} \mathrm{adj}\big(\tilde{\mathbf{G}}(z)\big) &= \big|\tilde{\mathbf{G}}(z)\big| \cdot \tilde{\mathbf{G}}(z)^{-1} \\ &= \begin{pmatrix} |\mathbf{G}(z)| \cdot \mathbf{I}_{\mu\times\mu} + \tilde{\mathbf{C}} \cdot \mathrm{adj}(\mathbf{Y}(z)) \cdot \tilde{\mathbf{C}}^\dagger & -\tilde{\mathbf{C}} \cdot \mathrm{adj}(\mathbf{Y}(z)) \\ -\mathrm{adj}(\mathbf{Y}(z)) \cdot \tilde{\mathbf{C}}^\dagger & \mathrm{adj}(\mathbf{Y}(z)) \end{pmatrix}, \end{aligned} \tag{77}$$

where

$$\tilde{\mathbf{G}}(z) = \begin{pmatrix} \mathbf{I}_{\mu\times\mu} & \tilde{\mathbf{C}} \\ \tilde{\mathbf{C}}^\dagger & z\mathbf{I}_{\nu\times\nu} \end{pmatrix}, \quad \mathbf{Y}(z) := z \cdot \mathbf{I}_{\nu\times\nu} - \tilde{\mathbf{C}}^\dagger \tilde{\mathbf{C}}, \quad |\mathbf{Y}(z)| = \big|\tilde{\mathbf{G}}(z)\big| = |\mathbf{G}(z)|, \tag{78}$$

and the following linear algebra formula have been used:

$$\begin{pmatrix} \mathbf{A}_{n\times n} & \mathbf{B}_{n\times m} \\ \mathbf{C}_{m\times n} & \mathbf{D}_{m\times m} \end{pmatrix}^{-1} = \begin{pmatrix} \mathbf{A}^{-1} + \mathbf{A}^{-1}\mathbf{B}\mathbf{E}^{-1}\mathbf{C}\mathbf{A}^{-1} & -\mathbf{A}^{-1}\mathbf{B}\mathbf{E}^{-1} \\ -\mathbf{E}^{-1}\mathbf{C}\mathbf{A}^{-1} & \mathbf{E}^{-1} \end{pmatrix}, \tag{79}$$

with $\mathbf{E} := \mathbf{D} - \mathbf{C}\mathbf{A}^{-1}\mathbf{B}$. Moreover, we have

$$\mathbf{Y}(z) = \begin{pmatrix} z - \gamma_1^*\gamma_1 & 0 & \cdots & 0 \\ 0 & z - \gamma_2^*\gamma_2 & & \vdots \\ \vdots & & \ddots & 0 \\ 0 & \cdots & 0 & z - \gamma_\nu^*\gamma_\nu \end{pmatrix} = \begin{pmatrix} z - \lambda_1 & 0 & \cdots & 0 \\ 0 & z - \lambda_2 & & \vdots \\ \vdots & & \ddots & 0 \\ 0 & \cdots & 0 & z - \lambda_\nu \end{pmatrix}, \tag{80}$$

where $\lambda_i := \gamma_i^*\gamma_i$. Hence, we obtain

$$\mathrm{adj}(\mathbf{Y}(z)) = |\mathbf{Y}(z)| \cdot \mathbf{Y}(z)^{-1}$$

$$= \prod_{i=1}^{\nu}(z-\lambda_i) \cdot \begin{pmatrix} (z-\lambda_1)^{-1} & 0 & \cdots & 0 \\ 0 & (z-\lambda_2)^{-1} & & \vdots \\ \vdots & & \ddots & 0 \\ 0 & \cdots & 0 & (z-\lambda_\nu)^{-1} \end{pmatrix}$$

$$= \begin{pmatrix} \prod_{i=2}^{\nu}(z-\lambda_i) & 0 & \cdots & 0 \\ 0 & \prod_{\substack{i=1\\(i\neq 2)}}^{\nu}(z-\lambda_i) & & \vdots \\ \vdots & & \ddots & 0 \\ 0 & \cdots & 0 & \prod_{i=1}^{\nu-1}(z-\lambda_i) \end{pmatrix} = \begin{pmatrix} \mathcal{G}_1(z) & 0 & \cdots & 0 \\ 0 & \mathcal{G}_2(z) & & \vdots \\ \vdots & & \ddots & 0 \\ 0 & \cdots & 0 & \mathcal{G}_\nu(z) \end{pmatrix}, \quad (81)$$

where

$$\mathcal{G}_q(z) := \prod_{\substack{p=1\\(p\neq q)}}^{\nu}(z-\lambda_p). \tag{82}$$

Substituting Eq. (81) into Eq. (77), we finally obtain

$$\mathrm{adj}(\tilde{\mathbf{G}}(z)) = \begin{pmatrix} z\mathcal{G}_1(z) & & & & & \mathbf{0} & -\gamma_1\mathcal{G}_1(z) & 0 & \cdots & 0 \\ & z\mathcal{G}_2(z) & & & & & 0 & -\gamma_2\mathcal{G}_2(z) & & \vdots \\ & & \ddots & & & & \vdots & & \ddots & 0 \\ & & & z\mathcal{G}_\nu(z) & & & 0 & \cdots & 0 & -\gamma_\nu\mathcal{G}_\nu(z) \\ & & & & \mathcal{G}(z) & & 0 & 0 & \cdots & 0 \\ \mathbf{0} & & & & & \mathcal{G}(z) & 0 & 0 & \cdots & 0 \\ -\gamma_1^*\mathcal{G}_1(z) & 0 & \cdots & 0 & 0 & 0 & \mathcal{G}_1(z) & & & \mathbf{0} \\ 0 & -\gamma_2^*\mathcal{G}_2(z) & & \vdots & 0 & 0 & & \mathcal{G}_2(z) & & \\ \vdots & & \ddots & 0 & \vdots & \vdots & & & \ddots & \\ 0 & \cdots & 0 & -\gamma_\nu^*\mathcal{G}_\nu(z) & 0 & 0 & \mathbf{0} & & & \mathcal{G}_\nu(z) \end{pmatrix},$$

(83)

where the following relations have been used:

$$\mathcal{G}(z) = |\mathbf{G}(z)| = \prod_{p=1}^{\nu}(z-\lambda_p) = (z-\lambda_q)\mathcal{G}_q(z). \tag{84}$$

## 6. Explicit expressions of $\mathcal{F}(z)$ and $\mathcal{V}(z)$

In this section, we derive the explicit expressions of $\mathcal{F}(z)$ and $\mathcal{V}(z)$ by using Eq. (83). First, we calculate $\mathcal{F}(z)$ in Eq. (75) as follows:

$$\begin{aligned}
\mathcal{F}(z) &= \sum_{p=1}^{n}\sum_{q=1}^{n}\left[h_{pq}\cdot\mathrm{adj}(\tilde{\mathbf{G}})_{qp}\right] \\
&= \left(\sum_{p=1}^{\mu}\sum_{q=1}^{\mu}+\sum_{p=1}^{\mu}\sum_{q=\mu+1}^{n}+\sum_{p=\mu+1}^{n}\sum_{q=1}^{\mu}+\sum_{p=\mu+1}^{n}\sum_{q=\mu+1}^{n}\right)\left[h_{pq}\cdot\mathrm{adj}(\tilde{\mathbf{G}})_{qp}\right] \\
&= \sum_{i_1=1}^{\mu}\sum_{i_2=1}^{\mu}\left[h_{i_1 i_2}^{\alpha\alpha}\cdot\mathrm{adj}(\tilde{\mathbf{G}})_{i_2 i_1}^{\alpha\alpha}\right]+\sum_{i=1}^{\mu}\sum_{j=1}^{\nu}\left[h_{ij}^{\alpha\beta}\cdot\mathrm{adj}(\tilde{\mathbf{G}})_{ji}^{\beta\alpha}\right]+\sum_{j=1}^{\nu}\sum_{i=1}^{\mu}\left[h_{ji}^{\beta\alpha}\cdot\mathrm{adj}(\tilde{\mathbf{G}})_{ij}^{\alpha\beta}\right]+\sum_{j_1=1}^{\nu}\sum_{j_2=1}^{\nu}\left[h_{j_1 j_2}^{\beta\beta}\cdot\mathrm{adj}(\tilde{\mathbf{G}})_{j_2 j_1}^{\beta\beta}\right] \\
&= \sum_{i=1}^{\mu}\left[h_{ii}^{\alpha\alpha}\cdot\mathrm{adj}(\tilde{\mathbf{G}})_{ii}^{\alpha\alpha}\right]+\sum_{j=1}^{\nu}\left[h_{jj}^{\alpha\beta}\cdot\mathrm{adj}(\tilde{\mathbf{G}})_{jj}^{\beta\alpha}\right]+\sum_{j=1}^{\nu}\left[h_{jj}^{\beta\alpha}\cdot\mathrm{adj}(\tilde{\mathbf{G}})_{jj}^{\alpha\beta}\right]+\sum_{j=1}^{\nu}\left[h_{jj}^{\beta\beta}\cdot\mathrm{adj}(\tilde{\mathbf{G}})_{jj}^{\beta\beta}\right] \\
&= \sum_{i=1}^{\nu}\left[\langle\phi_i|\hat{h}|\phi_i\rangle\cdot z\mathcal{G}_i(z)\right]+\sum_{i=\nu+1}^{\mu}\left[\langle\phi_i|\hat{h}|\phi_i\rangle\cdot\mathcal{G}(z)\right] \\
&\quad -\sum_{j=1}^{\nu}\left[\langle\phi_j|\hat{h}|\theta_j\rangle\cdot\gamma_j^*\mathcal{G}_j(z)\right]-\sum_{j=1}^{\nu}\left[\langle\theta_j|\hat{h}|\phi_j\rangle\cdot\gamma_j\mathcal{G}_j(z)\right]+\sum_{j=1}^{\nu}\left[z\langle\theta_j|\hat{h}|\theta_j\rangle\cdot\mathcal{G}_j(z)\right] \\
&= \sum_{i=1}^{\nu}\left\{z\left(\langle\phi_i|\hat{h}|\phi_i\rangle+\langle\theta_i|\hat{h}|\theta_i\rangle\right)-2\,\mathrm{Re}\left[\gamma_i\langle\theta_i|\hat{h}|\phi_i\rangle\right]\right\}\mathcal{G}_i(z)+\mathcal{G}(z)\sum_{i=\nu+1}^{\mu}\langle\phi_i|\hat{h}|\phi_i\rangle, \qquad (85)
\end{aligned}$$

where we have decomposed $\mathrm{adj}(\tilde{\mathbf{G}}(z))$ into the following four block matrices:

$$\mathrm{adj}(\tilde{\mathbf{G}}(z))_{n\times n} = \begin{pmatrix} \mathrm{adj}(\tilde{\mathbf{G}}(z))_{\mu\times\mu}^{\alpha\alpha} & \mathrm{adj}(\tilde{\mathbf{G}}(z))_{\nu\times\mu}^{\alpha\beta} \\ \mathrm{adj}(\tilde{\mathbf{G}}(z))_{\mu\times\nu}^{\beta\alpha} & \mathrm{adj}(\tilde{\mathbf{G}}(z))_{\nu\times\nu}^{\beta\beta} \end{pmatrix}. \qquad (86)$$

Next, we calculate $\mathcal{V}(z)$ in Eq. (75) as follows:

$$\begin{aligned}
\mathcal{V}(z) &= \frac{1}{4}\sum_{i=1}^{n}\sum_{j=1}^{n}\sum_{k=1}^{n}\sum_{l=1}^{n}V_{ij;kl}\cdot\mathrm{adj}^{(2)}(\tilde{\mathbf{G}})_{kl;ij} \\
&= \frac{1}{4}\sum_{i=1}^{n}\sum_{j=1}^{n}\sum_{k=1}^{n}\sum_{l=1}^{n}V_{ij;kl}\cdot\frac{1}{|\tilde{\mathbf{G}}|}\left[\mathrm{adj}(\tilde{\mathbf{G}})_{ki}\mathrm{adj}(\tilde{\mathbf{G}})_{lj}-\mathrm{adj}(\tilde{\mathbf{G}})_{kj}\mathrm{adj}(\tilde{\mathbf{G}})_{li}\right] \\
&= \frac{1}{4}\sum_{i=1}^{n}\sum_{j=1}^{n}\sum_{k=1}^{n}\sum_{l=1}^{n}\frac{1}{|\tilde{\mathbf{G}}|}\left[V_{ij;kl}\cdot\mathrm{adj}(\tilde{\mathbf{G}})_{ik}^{*}\mathrm{adj}(\tilde{\mathbf{G}})_{jl}^{*}+V_{ji;kl}\cdot\mathrm{adj}(\tilde{\mathbf{G}})_{jk}^{*}\mathrm{adj}(\tilde{\mathbf{G}})_{il}^{*}\right] \\
&= \frac{1}{2}\sum_{i=1}^{n}\sum_{\substack{j=1\\(j\neq i)}}^{n}\sum_{k=1}^{n}\sum_{\substack{l=1\\(l\neq k)}}^{n}V_{ij;kl}\cdot\frac{1}{\mathcal{G}(z)}\mathrm{adj}(\tilde{\mathbf{G}})_{ik}^{*}\mathrm{adj}(\tilde{\mathbf{G}})_{jl}^{*}, \qquad (87)
\end{aligned}$$

where we have utilized the fact that $\mathrm{adj}^{(2)}(\tilde{\mathbf{G}})$ is related to $\mathrm{adj}(\tilde{\mathbf{G}})$ through a Jacobi identity [46,47]. Since $\mathrm{adj}^{(2)}(\tilde{\mathbf{G}})_{kl;ij}=0$ when $k=l$ or $i=j$, we have added the restrictions ($k\neq l$ and $i\neq j$) to the summations in Eq. (87). Then, substituting

$$\sum_{\substack{i=1 \\ (j \neq i)}}^{n} \sum_{j=1}^{n} \sum_{\substack{k=1 \\ (l \neq k)}}^{n} \sum_{l=1}^{n} = \sum_{i_1=1}^{\mu} \sum_{\substack{i_2=1 \\ (i_2 \neq i_1)}}^{\mu} \sum_{i_3=1}^{\mu} \sum_{\substack{i_4=1 \\ (i_4 \neq i_3)}}^{\mu} + \sum_{i_1=1}^{\mu} \sum_{\substack{i_2=1 \\ (i_2 \neq i_1)}}^{\mu} \sum_{i_3=1}^{\mu} \sum_{j_1=1}^{\nu} + \sum_{i_1=1}^{\mu} \sum_{\substack{i_2=1 \\ (i_2 \neq i_1)}}^{\mu} \sum_{j_1=1}^{\nu} \sum_{i_3=1}^{\mu} + \sum_{i_1=1}^{\mu} \sum_{\substack{i_2=1 \\ (i_2 \neq i_1)}}^{\mu} \sum_{j_1=1}^{\nu} \sum_{\substack{j_2=1 \\ (j_2 \neq j_1)}}^{\nu}$$

$$+ \sum_{i_1=1}^{\mu} \sum_{j_1=1}^{\nu} \sum_{i_2=1}^{\mu} \sum_{\substack{i_3=1 \\ (i_3 \neq i_2)}}^{\mu} + \sum_{i_1=1}^{\mu} \sum_{j_1=1}^{\nu} \sum_{i_2=1}^{\nu} \sum_{j_2=1}^{\mu} + \sum_{i_1=1}^{\mu} \sum_{j_1=1}^{\nu} \sum_{j_2=1}^{\nu} \sum_{i_2=1}^{\mu} + \sum_{i_1=1}^{\mu} \sum_{j_1=1}^{\nu} \sum_{j_2=1}^{\nu} \sum_{\substack{j_3=1 \\ (j_3 \neq j_2)}}^{\nu}$$

$$+ \sum_{j_1=1}^{\nu} \sum_{i_1=1}^{\mu} \sum_{i_2=1}^{\mu} \sum_{\substack{i_3=1 \\ (i_3 \neq i_2)}}^{\mu} + \sum_{j_1=1}^{\nu} \sum_{i_1=1}^{\mu} \sum_{i_2=1}^{\nu} \sum_{j_2=1}^{\mu} + \sum_{j_1=1}^{\nu} \sum_{i_1=1}^{\mu} \sum_{j_2=1}^{\nu} \sum_{i_2=1}^{\mu} + \sum_{j_1=1}^{\nu} \sum_{i_1=1}^{\mu} \sum_{j_2=1}^{\nu} \sum_{\substack{j_3=1 \\ (j_3 \neq j_2)}}^{\nu}$$

$$+ \sum_{j_1=1}^{\nu} \sum_{\substack{j_2=1 \\ (j_2 \neq j_1)}}^{\nu} \sum_{i_1=1}^{\mu} \sum_{\substack{i_2=1 \\ (i_2 \neq i_1)}}^{\mu} + \sum_{j_1=1}^{\nu} \sum_{\substack{j_2=1 \\ (j_2 \neq j_1)}}^{\nu} \sum_{i_1=1}^{\mu} \sum_{j_3=1}^{\nu} + \sum_{j_1=1}^{\nu} \sum_{\substack{j_2=1 \\ (j_2 \neq j_1)}}^{\nu} \sum_{j_3=1}^{\nu} \sum_{i_1=1}^{\mu} + \sum_{j_1=1}^{\nu} \sum_{\substack{j_2=1 \\ (j_2 \neq j_1)}}^{\nu} \sum_{j_3=1}^{\nu} \sum_{\substack{j_4=1 \\ (j_4 \neq j_3)}}^{\nu} , \quad (88)$$

into Eq. (87), we can decompose $\mathcal{V}(z)$ into 16 elements corresponding to the 16 block matrices of $\mathbf{V}(z)$:

$$\mathcal{V}(z) = v_{11} + v_{12} + v_{13} + v_{14}$$
$$+ v_{21} + v_{22} + v_{23} + v_{24}$$
$$+ v_{31} + v_{32} + v_{33} + v_{34}$$
$$+ v_{41} + v_{42} + v_{43} + v_{44} , \quad (89)$$

where we note that the block matrices of $\text{adj}(\tilde{\mathbf{G}}(z))$ have been required to obtain Eq. (89). For example, we have

$$v_{11} = \frac{1}{2} \sum_{i_1=1}^{\mu} \sum_{\substack{i_2=1 \\ (i_2 \neq i_1)}}^{\mu} \sum_{i_3=1}^{\mu} \sum_{\substack{i_4=1 \\ (i_4 \neq i_3)}}^{\mu} \langle \phi_{i_1} \phi_{i_2} \| \phi_{i_3} \phi_{i_4} \rangle \frac{\text{adj}(\tilde{\mathbf{G}})_{i_1 i_3}^{\alpha\alpha*} \text{adj}(\tilde{\mathbf{G}})_{i_2 i_4}^{\alpha\alpha*}}{\mathcal{G}(z)} , \quad (90)$$

$$v_{32} = \frac{1}{2} \sum_{j_1=1}^{\nu} \sum_{i_1=1}^{\mu} \sum_{i_2=1}^{\mu} \sum_{j_2=1}^{\nu} \langle \bar{\theta}_{j_1} \phi_{i_1} \| \phi_{i_2} \bar{\theta}_{j_2} \rangle \frac{\text{adj}(\tilde{\mathbf{G}})_{j_1 i_2}^{\beta\alpha*} \text{adj}(\tilde{\mathbf{G}})_{i_1 j_2}^{\alpha\beta*}}{\mathcal{G}(z)} , \quad (91)$$

with the definitions of Eq. (97). Since the four block matrices of $\text{adj}(\tilde{\mathbf{G}}(z))$ are diagonal matrices, after some algebra we have

$$v_{11} + v_{44} = \frac{1}{2} \sum_{k=1}^{\nu} \sum_{\substack{l=1 \\ (l \neq k)}}^{\nu} \left( z^2 \langle \phi_k \phi_l \| \phi_k \phi_l \rangle + \langle \bar{\theta}_k \bar{\theta}_l \| \bar{\theta}_k \bar{\theta}_l \rangle \right) \mathcal{G}_{k,l}(z) + \sum_{k=1}^{\nu} \sum_{l=\nu+1}^{\mu} z \langle \phi_k \phi_l \| \phi_k \phi_l \rangle \mathcal{G}_k(z)$$
$$+ \frac{1}{2} \sum_{k=\nu+1}^{\mu} \sum_{\substack{l=\nu+1 \\ (l \neq k)}}^{\mu} \langle \phi_k \phi_l \| \phi_k \phi_l \rangle \mathcal{G}(z) , \quad (92)$$

$$v_{22} + v_{33} + v_{23} + v_{32} = \sum_{k=1}^{\nu} \sum_{\substack{l=1 \\ (l \neq k)}}^{\nu} \left( z \langle \phi_k \bar{\theta}_l \| \phi_k \bar{\theta}_l \rangle + \gamma_k \gamma_l^* \langle \bar{\theta}_k \phi_l \| \phi_k \bar{\theta}_l \rangle \right) \mathcal{G}_{k,l}(z)$$
$$+ \sum_{k=1}^{\nu} \sum_{l=\nu+1}^{\mu} \langle \bar{\theta}_k \phi_l \| \bar{\theta}_k \phi_l \rangle \mathcal{G}_k(z) , \quad (93)$$

$$v_{12} + v_{13} + v_{21} + v_{31} = \sum_{k=1}^{v} \sum_{\substack{l=1 \\ (l \neq k)}}^{v} \langle \phi_k \phi_l \| \phi_k \theta_l \rangle z(-\gamma_l^*) \mathcal{G}_{k,l}(z) + \sum_{k=1}^{v} \sum_{l=v+1}^{\mu} \langle \phi_k \phi_l \| \theta_k \phi_l \rangle (-\gamma_k^*) \mathcal{G}_k(z)$$
$$+ \sum_{k=1}^{v} \sum_{\substack{l=1 \\ (l \neq k)}}^{v} \langle \phi_k \phi_l \| \phi_k \theta_l \rangle^* z(-\gamma_l) \mathcal{G}_{k,l}(z) + \sum_{k=1}^{v} \sum_{l=v+1}^{\mu} \langle \phi_k \phi_l \| \theta_k \phi_l \rangle^* (-\gamma_k) \mathcal{G}_k(z), \quad (94)$$

$$v_{14} + v_{41} = \frac{1}{2} \sum_{k=1}^{v} \sum_{\substack{l=1 \\ (l \neq k)}}^{v} \langle \phi_k \phi_l \| \theta_k \theta_l \rangle \gamma_k^* \gamma_l^* \mathcal{G}_{k,l}(z) + \frac{1}{2} \sum_{k=1}^{v} \sum_{\substack{l=1 \\ (l \neq k)}}^{v} \langle \phi_k \phi_l \| \theta_k \theta_l \rangle^* \gamma_k \gamma_l \mathcal{G}_{k,l}(z), \quad (95)$$

$$v_{24} + v_{34} + v_{42} + v_{43} = \sum_{k=1}^{v} \sum_{\substack{l=1 \\ (l \neq k)}}^{v} \langle \overline{\theta}_k \phi_l \| \overline{\theta}_k \overline{\theta}_l \rangle (-\gamma_l^*) \mathcal{G}_{k,l}(z) + \sum_{k=1}^{v} \sum_{\substack{l=1 \\ (l \neq k)}}^{v} \langle \overline{\theta}_k \phi_l \| \overline{\theta}_k \overline{\theta}_l \rangle^* (-\gamma_l) \mathcal{G}_{k,l}(z), \quad (96)$$

where

$$\langle \phi_k \overline{\theta}_l \| \phi_k \overline{\theta}_l \rangle := z \langle \phi_k \theta_l | \phi_k \theta_l \rangle - \langle \phi_k \theta_l | \theta_l \phi_k \rangle,$$
$$\langle \overline{\theta}_k \phi_l \| \phi_k \overline{\theta}_l \rangle := \langle \theta_k \phi_l | \phi_k \theta_l \rangle - z \langle \theta_k \phi_l | \theta_l \phi_k \rangle,$$
$$\langle \overline{\theta}_k \phi_l \| \overline{\theta}_k \phi_l \rangle := z \langle \theta_k \phi_l | \theta_k \phi_l \rangle - \langle \theta_k \phi_l | \phi_l \theta_k \rangle, \quad (97)$$
$$\langle \overline{\theta}_k \phi_l \| \overline{\theta}_k \overline{\theta}_l \rangle := z \langle \theta_k \phi_l | \theta_k \theta_l \rangle - \langle \theta_k \phi_l | \theta_l \theta_k \rangle,$$
$$\langle \overline{\theta}_k \overline{\theta}_l \| \overline{\theta}_k \overline{\theta}_l \rangle := z^2 \langle \theta_k \theta_l | \theta_k \theta_l \rangle - z^2 \langle \theta_k \theta_l | \theta_l \theta_k \rangle,$$

and

$$\mathcal{G}_{q,r}(z) := \prod_{\substack{p=1 \\ (p \neq q,r)}}^{v} (z - \lambda_p) = \frac{\mathcal{G}_q(z) \cdot \mathcal{G}_r(z)}{\mathcal{G}(z)} \quad (q \neq r). \quad (98)$$

After some algebra we finally obtain

$$\mathcal{V}(z) = \frac{1}{2} \mathcal{G}(z) \sum_{k=v+1}^{\mu} \sum_{\substack{l=v+1 \\ (l \neq k)}}^{\mu} \langle \phi_k \phi_l \| \phi_k \phi_l \rangle$$
$$+ \sum_{k=1}^{v} \sum_{\substack{l=1 \\ (l \neq k)}}^{v} \left\{ z^2 \left( \frac{1}{2} \langle \phi_k \phi_l \| \phi_k \phi_l \rangle + \langle \phi_k \theta_l | \phi_k \theta_l \rangle + \frac{1}{2} \langle \theta_k \theta_l \| \theta_k \theta_l \rangle \right) - z \langle \phi_k \theta_l | \theta_l \phi_k \rangle \right.$$
$$\left. + \mathrm{Re} \left[ \gamma_k \gamma_l \langle \phi_k \phi_l \| \theta_k \theta_l \rangle - 2 z \gamma_l \left( \langle \phi_k \phi_l \| \phi_k \theta_l \rangle + \langle \phi_k \theta_l \| \theta_k \theta_l \rangle \right) \right] - \gamma_k \gamma_l^* \left( z \langle \theta_k \phi_l | \theta_l \phi_k \rangle - \langle \theta_k \phi_l | \phi_k \theta_l \rangle \right) \right\} \mathcal{G}_{k,l}(z)$$
$$+ \sum_{k=1}^{v} \sum_{l=v+1}^{\mu} \left\{ z \left( \langle \phi_k \phi_l \| \phi_k \phi_l \rangle + \langle \theta_k \phi_l | \theta_k \phi_l \rangle \right) - \langle \theta_k \phi_l | \phi_l \theta_k \rangle - 2 \mathrm{Re} \left[ \gamma_k \langle \phi_k \phi_l \| \theta_k \phi_l \rangle \right] \right\} \mathcal{G}_k(z). \quad (99)$$

We here note that, strictly speaking, only for closed-shell molecules Eqs. (85) and (99) should hold, because we do not adopt the CO model.

## 7. Common orbital model

Following Pons Viver's proposals [22], we introduce the CO model as follows:

$\alpha$ spatial orbitals : $\vartheta_1, \cdots, \vartheta_{n_c}, \phi_{n_c+1}, \cdots, \phi_\mu,$
$\beta$ spatial orbitals : $\vartheta_1, \cdots, \vartheta_{n_c}, \theta_{n_c+1}, \cdots, \theta_\nu,$ \quad (100)

where $\vartheta_i$ are spatial COs and $n_c$ is the number of the COs. Since the overlap between $\alpha$ and $\beta$ spatial orbitals is 1 or 0 in the CO model [22], we have

$$\tilde{\mathbf{C}} = \begin{pmatrix} \langle\phi_1|\theta_1\rangle & \langle\phi_1|\theta_2\rangle & \cdots & \langle\phi_1|\theta_\nu\rangle \\ \langle\phi_2|\theta_1\rangle & \langle\phi_2|\theta_2\rangle & \cdots & \langle\phi_2|\theta_\nu\rangle \\ \langle\phi_3|\theta_1\rangle & \langle\phi_3|\theta_2\rangle & \cdots & \langle\phi_3|\theta_\nu\rangle \\ \vdots & \vdots & & \vdots \\ \vdots & \vdots & & \vdots \\ \langle\phi_\mu|\theta_1\rangle & \langle\phi_\mu|\theta_2\rangle & \cdots & \langle\phi_\mu|\theta_\nu\rangle \end{pmatrix} = \begin{pmatrix} \overbrace{\begin{matrix}1 & & & \\ & 1 & & \\ & & \ddots & \\ & & & 1 \\ 0 & & & \end{matrix}}^{n_c} & \overbrace{\begin{matrix}\mathbf{0} & 0 & 0 \\ & 0 & 0 \\ & \vdots & \vdots \\ & \vdots & \vdots \\ 1 & 0 & 0\end{matrix}}^{\nu-n_c} \\ 0 \cdots \cdots \cdots 0\ 0\ 0 \\ \vdots \quad\quad \vdots\ \vdots\ \vdots \\ 0 \cdots \cdots \cdots 0\ 0\ 0 \end{pmatrix} \left.\begin{matrix}\\ \\ \\ \\ \\ \end{matrix}\right\}n_c \left.\begin{matrix}\\ \\ \\\end{matrix}\right\}\mu-n_c \qquad (101)$$

Calculating $\mathrm{adj}(\tilde{\mathbf{G}}(z))$ of Eq. (77) by using $\tilde{\mathbf{C}}$ of Eq. (101), we obtain

$$\mathrm{adj}(\tilde{\mathbf{G}}(z)) =$$

$$\begin{pmatrix} z\mathcal{R}(z) & & & & & & \mathbf{0} & -\mathcal{R}(z) & 0 & \cdots & \cdots & 0 \\ & \ddots & & & & & & 0 & \ddots & & & \vdots \\ & & z\mathcal{R}(z) & & & & & \vdots & & -\mathcal{R}(z) & & \vdots \\ & & & \mathcal{G}(z) & & & & \vdots & & & 0 & \vdots \\ & & & & \mathcal{G}(z) & & & 0 & \cdots & \cdots & \cdots & 0 \\ & & & & & \mathcal{G}(z) & & 0 & 0 & \cdots & \cdots & 0 \\ \mathbf{0} & & & & & & \mathcal{G}(z) & 0 & 0 & \cdots & \cdots & 0 \\ -\mathcal{R}(z) & 0 & \cdots & \cdots & 0 & 0 & 0 & \mathcal{R}(z) & & & & \mathbf{0} \\ 0 & \ddots & & & \vdots & 0 & 0 & & \ddots & & & \\ \vdots & & -\mathcal{R}(z) & & \vdots & \vdots & \vdots & & & \mathcal{R}(z) & & \\ \vdots & & & 0 & \vdots & \vdots & \vdots & & & & \mathcal{G}(z)/z & \\ 0 & \cdots & \cdots & \cdots & 0 & 0 & 0 & \mathbf{0} & & & & \mathcal{G}(z)/z \end{pmatrix} \begin{matrix}\left.\begin{matrix}\\ \\ \\\end{matrix}\right\}n_c \\ \left.\begin{matrix}\\ \\ \\\end{matrix}\right\}\nu-n_c \\ \left.\begin{matrix}\\ \\ \\\end{matrix}\right\}\mu-\nu, \\ \left.\begin{matrix}\\ \\ \\\end{matrix}\right\}n_c \\ \left.\begin{matrix}\\ \\ \\\end{matrix}\right\}\nu-n_c \end{matrix}$$

(102)

where

$$\mathcal{G}(z) = (z-1)^{n_c} z^{\nu-n_c}, \qquad \mathcal{R}(z) = (z-1)^{n_c-1} z^{\nu-n_c} = \frac{\mathcal{G}(z)}{z-1}. \qquad (103)$$

Next, calculating $\mathcal{F}(z)$ of Eq. (85) by using both $\mathrm{adj}(\tilde{\mathbf{G}}(z))$ of Eq. (102) and $\mathcal{G}(z)$ of Eq. (103), we obtain

$$\mathcal{F}(z) = \left(\sum_{i=1}^{n_c} + \sum_{i=n_c+1}^{\mu}\right)\left[h_{ii}^{\alpha\alpha} \cdot \mathrm{adj}\left(\tilde{\mathbf{G}}\right)_{ii}^{\alpha\alpha}\right] + \sum_{j=1}^{\nu}\left[h_{jj}^{\alpha\beta} \cdot \mathrm{adj}\left(\tilde{\mathbf{G}}\right)_{jj}^{\beta\alpha}\right]$$

$$+ \sum_{j=1}^{\nu}\left[h_{jj}^{\beta\alpha} \cdot \mathrm{adj}\left(\tilde{\mathbf{G}}\right)_{jj}^{\alpha\beta}\right] + \left(\sum_{j=1}^{n_c} + \sum_{j=n_c+1}^{\nu}\right)\left[h_{jj}^{\beta\beta} \cdot \mathrm{adj}\left(\tilde{\mathbf{G}}\right)_{jj}^{\beta\beta}\right]$$

$$= \sum_{i=1}^{n_c}\langle\phi_i|\hat{h}|\phi_i\rangle\frac{z}{z-1}\mathcal{G}(z) + \sum_{i=n_c+1}^{\mu}\langle\phi_i|\hat{h}|\phi_i\rangle\mathcal{G}(z)$$

$$- \sum_{j=1}^{n_c}\langle\phi_j|\hat{h}|\theta_j\rangle\frac{1}{z-1}\mathcal{G}(z) - \sum_{j=1}^{n_c}\langle\theta_j|\hat{h}|\phi_j\rangle\frac{1}{z-1}\mathcal{G}(z)$$

$$+ \sum_{j=1}^{n_c}\langle\theta_j|\hat{h}|\theta_j\rangle\frac{z}{z-1}\mathcal{G}(z) + \sum_{j=n_c+1}^{\nu}\langle\theta_j|\hat{h}|\theta_j\rangle\mathcal{G}(z)$$

$$= \sum_{i=1}^{n_c}\langle\phi_i|\hat{h}|\phi_i\rangle\frac{z}{z-1}\mathcal{G}(z) + \sum_{i=1}^{n_c}\langle\phi_i|\hat{h}|\phi_i\rangle\frac{-1}{z-1}\mathcal{G}(z) + \sum_{i=n_c+1}^{\mu}\langle\phi_i|\hat{h}|\phi_i\rangle\mathcal{G}(z)$$

$$+ \sum_{j=1}^{n_c}\langle\theta_j|\hat{h}|\theta_j\rangle\frac{z}{z-1}\mathcal{G}(z) + \sum_{j=1}^{n_c}\langle\theta_j|\hat{h}|\theta_j\rangle\frac{-1}{z-1}\mathcal{G}(z) + \sum_{j=n_c+1}^{\nu}\langle\theta_j|\hat{h}|\theta_j\rangle\mathcal{G}(z)$$

$$= \left(\sum_{i=1}^{\mu}\langle\phi_i|\hat{h}|\phi_i\rangle + \sum_{j=1}^{\nu}\langle\theta_j|\hat{h}|\theta_j\rangle\right)\mathcal{G}(z). \tag{104}$$

Moreover, calculating $\mathcal{V}(z)$ of Eq. (87) by using both $\mathrm{adj}(\tilde{\mathbf{G}}(z))$ of Eq. (102) and $\mathcal{G}(z)$ of Eq. (103), after some algebra, like Eq. (89), we have

$$\mathcal{V}(z) = w_{11} + w_{12} + w_{13} + w_{14}$$
$$+ w_{21} + w_{22} + w_{23} + w_{24}$$
$$+ w_{31} + w_{32} + w_{33} + w_{34}$$
$$+ w_{41} + w_{42} + w_{43} + w_{44}, \tag{105}$$

where

$$w_{11} + w_{44} = \frac{1}{2}\sum_{k=1}^{n_c}\sum_{l=1}^{n_c}\langle\phi_k\phi_l\|\phi_k\phi_l\rangle\frac{z^2\mathcal{G}(z)}{(z-1)^2} + \frac{1}{2}\sum_{k=1}^{n_c}\sum_{l=n_c+1}^{\mu}\langle\phi_k\phi_l\|\phi_k\phi_l\rangle\frac{z\mathcal{G}(z)}{z-1}$$

$$+ \frac{1}{2}\sum_{k=n_c+1}^{\mu}\sum_{l=1}^{n_c}\langle\phi_k\phi_l\|\phi_k\phi_l\rangle\frac{z\mathcal{G}(z)}{z-1} + \frac{1}{2}\sum_{k=n_c+1}^{\mu}\sum_{l=n_c+1}^{\mu}\langle\phi_k\phi_l\|\phi_k\phi_l\rangle\mathcal{G}(z)$$

$$+ \frac{1}{2}\sum_{k=1}^{n_c}\sum_{l=1}^{n_c}\langle\bar{\theta}_k\bar{\theta}_l\|\bar{\theta}_k\bar{\theta}_l\rangle\frac{\mathcal{G}(z)}{(z-1)^2} + \frac{1}{2}\sum_{k=1}^{n_c}\sum_{l=n_c+1}^{\nu}\langle\bar{\theta}_k\bar{\theta}_l\|\bar{\theta}_k\bar{\theta}_l\rangle\frac{\mathcal{G}(z)}{z(z-1)}$$

$$+ \frac{1}{2}\sum_{k=n_c+1}^{\nu}\sum_{l=1}^{n_c}\langle\bar{\theta}_k\bar{\theta}_l\|\bar{\theta}_k\bar{\theta}_l\rangle\frac{\mathcal{G}(z)}{z(z-1)} + \frac{1}{2}\sum_{k=n_c+1}^{\nu}\sum_{l=n_c+1}^{\nu}\langle\bar{\theta}_k\bar{\theta}_l\|\bar{\theta}_k\bar{\theta}_l\rangle\frac{\mathcal{G}(z)}{z^2}, \tag{106}$$

$$w_{22} + w_{33} + w_{23} + w_{32} = \sum_{k=1}^{n_c}\sum_{l=1}^{n_c}\langle\phi_k\bar{\theta}_l\|\phi_k\bar{\theta}_l\rangle\frac{z\mathcal{G}(z)}{(z-1)^2} + \sum_{k=1}^{n_c}\sum_{l=n_c+1}^{\nu}\langle\phi_k\bar{\theta}_l\|\phi_k\bar{\theta}_l\rangle\frac{\mathcal{G}(z)}{z-1}$$

$$+ \sum_{k=n_c+1}^{\mu}\sum_{l=1}^{n_c}\langle\phi_k\bar{\theta}_l\|\phi_k\bar{\theta}_l\rangle\frac{\mathcal{G}(z)}{z-1} + \sum_{k=n_c+1}^{\mu}\sum_{l=n_c+1}^{\nu}\langle\phi_k\bar{\theta}_l\|\phi_k\bar{\theta}_l\rangle\frac{\mathcal{G}(z)}{z}$$

$$+ \sum_{k=1}^{n_c}\sum_{l=1}^{n_c}\langle\bar{\theta}_k\phi_l\|\phi_k\bar{\theta}_l\rangle\frac{\mathcal{G}(z)}{(z-1)^2}, \tag{107}$$

$$w_{12} + w_{13} + w_{21} + w_{31} = -\sum_{k=1}^{n_c}\sum_{l=1}^{n_c}\langle\phi_k\phi_l \| \phi_k\theta_l\rangle \frac{z\mathcal{G}(z)}{(z-1)^2} - \sum_{k=n_c+1}^{\mu}\sum_{l=1}^{n_c}\langle\phi_k\phi_l \| \phi_k\theta_l\rangle \frac{\mathcal{G}(z)}{z-1}$$
$$-\sum_{k=1}^{n_c}\sum_{l=1}^{n_c}\langle\phi_k\phi_l \| \phi_k\theta_l\rangle^* \frac{z\mathcal{G}(z)}{(z-1)^2} - \sum_{k=n_c+1}^{\mu}\sum_{l=1}^{n_c}\langle\phi_k\phi_l \| \phi_k\theta_l\rangle^* \frac{\mathcal{G}(z)}{z-1}, \tag{108}$$

$$w_{14} + w_{41} = \frac{1}{2}\sum_{k=1}^{n_c}\sum_{l=1}^{n_c}\langle\phi_k\phi_l \| \theta_k\theta_l\rangle \frac{\mathcal{G}(z)}{(z-1)^2} + \frac{1}{2}\sum_{k=1}^{n_c}\sum_{l=1}^{n_c}\langle\phi_k\phi_l \| \theta_k\theta_l\rangle^* \frac{\mathcal{G}(z)}{(z-1)^2}, \tag{109}$$

$$w_{24} + w_{34} + w_{42} + w_{43} = -\sum_{k=1}^{n_c}\sum_{l=1}^{n_c}\langle\bar{\theta}_k\phi_l \| \bar{\theta}_k\bar{\theta}_l\rangle \frac{\mathcal{G}(z)}{(z-1)^2} - \sum_{k=n_c+1}^{\nu}\sum_{l=1}^{n_c}\langle\bar{\theta}_k\phi_l \| \bar{\theta}_k\bar{\theta}_l\rangle \frac{\mathcal{G}(z)}{z(z-1)}$$
$$-\sum_{k=1}^{n_c}\sum_{l=1}^{n_c}\langle\bar{\theta}_k\phi_l \| \bar{\theta}_k\bar{\theta}_l\rangle^* \frac{\mathcal{G}(z)}{(z-1)^2} - \sum_{k=n_c+1}^{\nu}\sum_{l=1}^{n_c}\langle\bar{\theta}_k\phi_l \| \bar{\theta}_k\bar{\theta}_l\rangle^* \frac{\mathcal{G}(z)}{z(z-1)}, \tag{110}$$

with the definitions of Eq. (97). After some lengthy algebra we finally obtain

$$\mathcal{V}(z) = \left(\frac{1}{2}\sum_{k=1}^{\mu}\sum_{l=1}^{\mu}\langle\phi_k\phi_l \| \phi_k\phi_l\rangle + \frac{1}{2}\sum_{k=1}^{\nu}\sum_{l=1}^{\nu}\langle\theta_k\theta_l \| \theta_k\theta_l\rangle + \sum_{k=1}^{\mu}\sum_{l=1}^{\nu}\langle\phi_k\theta_l | \phi_k\theta_l\rangle - \sum_{k=n_c+1}^{\mu}\sum_{l=n_c+1}^{\nu}\frac{1}{z}\langle\phi_k\theta_l | \theta_l\phi_k\rangle\right)\mathcal{G}(z).$$
$$\tag{111}$$

Here, according to Pons Viver's derivations [22], $1/z$ can be written as

$$\frac{1}{z} = -\frac{(S-|M|)(S+1+|M|)-\delta}{\delta(2|M|+\delta)}, \tag{112}$$

with $\delta = \nu - n_c$, whereas we have $z = C_{\nu-1}(S,M,N)$ [22]. In Eq. (111) the sum of the first, second, and third terms in the parenthesis is equal to the two-electron part of the UHF energy. We also note that, when by using both CO model and extended pairing theorem we obtain the corresponding orbitals satisfying Eq. (37), for both closed- and open-shell molecules Eqs. (104) and (111) hold.

## 8. EHF equations based on the CO model

Now we can express the EHF energy of Eq. (1) by using $\mathcal{G}(z)$, $\mathcal{F}(z)$, and $\mathcal{V}(z)$:

$$E = \frac{\langle\Psi|\hat{H}\hat{\mathcal{P}}_S|\Psi\rangle}{\langle\Psi|\hat{\mathcal{P}}_S|\Psi\rangle} = \frac{\mathcal{F}(z)+\mathcal{V}(z)}{\mathcal{G}(z)}. \tag{113}$$

Then, substituting Eqs. (104) and (111) into Eq. (113), we obtain the EHF energy based on the CO model:

$$E = \sum_{i=1}^{\mu}\langle\phi_i|\hat{h}|\phi_i\rangle + \sum_{j=1}^{\nu}\langle\theta_j|\hat{h}|\theta_j\rangle + \frac{1}{2}\sum_{k=1}^{\mu}\sum_{l=1}^{\mu}\langle\phi_k\phi_l \| \phi_k\phi_l\rangle + \frac{1}{2}\sum_{k=1}^{\nu}\sum_{l=1}^{\nu}\langle\theta_k\theta_l \| \theta_k\theta_l\rangle$$
$$+\sum_{k=1}^{\mu}\sum_{l=1}^{\nu}\langle\phi_k\theta_l | \phi_k\theta_l\rangle - \sum_{k=n_c+1}^{\mu}\sum_{l=n_c+1}^{\nu}\frac{1}{z}\langle\phi_k\theta_l | \theta_l\phi_k\rangle. \tag{114}$$

If we remove the last term in Eq. (114), the resulting equation is equal to the UHF energy. Interestingly, for closed-shell molecules Eq. (114) becomes the UHF energy when $n_c = \nu = \mu$, whereas even for closed-shell molecules Eq. (114) is not equivalent to the UHF energy when $n_c < \nu = \mu$.

Mayer et al. [10–12] derived the EHF equations based on the generalized Brillouin theorem instead of the Lagrangian multiplier technique. This is partially because the overlap integrals $\gamma_i = \langle \phi_i | \theta_i \rangle$ between the corresponding orbitals appear explicitly in the energy formula (please see Eqs. (85) and (99)) [12]. Conversely, since the EHF energy of Eq. (114) is quite similar to the UHF or ROHF energy and the Lagrangian multiplier technique has been applied to both UHF and ROHF formulations [16], we derive the EHF equations based on the Lagrangian multiplier technique in the present study. Incidentally, Mayer pointed out that the Lagrangian multiplier technique raises some questions of mathematical character, and he proposed the alternative method to the Lagrangian multiplier technique [16].

To obtain the EHF equations, we define the following Lagrangian $L$, like the UHF case:

$$L = \sum_{i=1}^{\mu} \langle \phi_i | \hat{h} | \phi_i \rangle + \sum_{j=1}^{\nu} \langle \theta_j | \hat{h} | \theta_j \rangle$$
$$+ \frac{1}{2} \sum_{i,j=1}^{\mu} \left( \langle \phi_i \phi_j | \phi_i \phi_j \rangle - \langle \phi_i \phi_j | \phi_j \phi_i \rangle \right) + \frac{1}{2} \sum_{i,j=1}^{\nu} \left( \langle \theta_i \theta_j | \theta_i \theta_j \rangle - \langle \theta_i \theta_j | \theta_j \theta_i \rangle \right) + \sum_{i=1}^{\mu} \sum_{j=1}^{\nu} \langle \phi_i \theta_j | \phi_i \theta_j \rangle$$
$$- \frac{1}{z} \sum_{i=n_c+1}^{\mu} \sum_{j=n_c+1}^{\nu} \langle \phi_i \theta_j | \theta_j \phi_i \rangle - \sum_{i,j=1}^{\mu} \varepsilon_{ji}^{\phi} \left( \langle \phi_i | \phi_j \rangle - \delta_{ij} \right) - \sum_{i,j=1}^{\nu} \varepsilon_{ji}^{\theta} \left( \langle \theta_i | \theta_j \rangle - \delta_{ij} \right), \qquad (115)$$

where $\varepsilon_{ji}^{\phi}$ and $\varepsilon_{ji}^{\theta}$ are Lagrangian multipliers. First, we require that the Lagrangian $L$ is stationary with respect to the variation of an orbital $\phi_k$ under the orthonormality constraint:

$$\delta L = \langle \delta \phi_k | \hat{F}_k^{\phi} | \phi_k \rangle - \langle \delta \phi_k | \sum_{j=1}^{\mu} \varepsilon_{jk}^{\phi} | \phi_j \rangle + \langle \phi_k | \hat{F}_k^{\phi} | \delta \phi_k \rangle - \sum_{j=1}^{\mu} \varepsilon_{kj}^{\phi} \langle \phi_j | \delta \phi_k \rangle = 0, \qquad (116)$$

where

$$\hat{F}_k^{\phi} := \hat{h} + \sum_{j=1}^{\mu} \left( \hat{J}_j^{\phi} - \hat{K}_j^{\phi} \right) + \sum_{j=1}^{\nu} \hat{J}_j^{\theta} - \frac{\eta_k^{\mu}}{z} \sum_{j=n_c+1}^{\nu} \hat{K}_j^{\theta}, \qquad (117)$$

$$\hat{J}_j^{\phi} := \langle \phi_j(2) | \frac{1}{r_{12}} | \phi_j(2) \rangle, \quad \hat{J}_j^{\theta} := \langle \theta_j(2) | \frac{1}{r_{12}} | \theta_j(2) \rangle, \qquad (118)$$

$$\hat{K}_j^{\phi} := \langle \phi_j(2) | \frac{\hat{P}_{12}}{r_{12}} | \phi_j(2) \rangle, \quad \hat{K}_j^{\theta} := \langle \theta_j(2) | \frac{\hat{P}_{12}}{r_{12}} | \theta_j(2) \rangle, \qquad (119)$$

$$\eta_k^{\mu} := \begin{cases} 1 & \text{if } n_c + 1 \leq k \leq \mu \\ 0 & \text{if } 1 \leq k \leq n_c \end{cases}, \qquad (120)$$

$\hat{P}_{12}$ is the permutation operator of electrons 1 and 2. Since $\delta \phi_k$ is arbitrary, therefore we obtain the following equations:

$$\hat{F}_k^\phi |\phi_k\rangle = \sum_{j=1}^{\mu} \varepsilon_{jk}^\phi |\phi_j\rangle, \tag{121}$$

$$\langle \phi_k | \hat{F}_k^\phi = \sum_{j=1}^{\mu} \varepsilon_{kj}^\phi \langle \phi_j |. \tag{122}$$

Similaly, requiring that the Lagrangian $L$ is stationary with respect to the variation of an orbital $\theta_k$, we finally obtain the following equations:

$$\hat{F}_k^\theta |\theta_k\rangle = \sum_{j=1}^{\nu} \varepsilon_{jk}^\theta |\theta_j\rangle, \tag{123}$$

$$\langle \theta_k | \hat{F}_k^\theta = \sum_{j=1}^{\nu} \varepsilon_{kj}^\theta \langle \theta_j |, \tag{124}$$

where

$$\hat{F}_k^\theta := \hat{h} + \sum_{j=1}^{\nu}\left(\hat{J}_j^\theta - \hat{K}_j^\theta\right) + \sum_{j=1}^{\mu}\hat{J}_j^\phi - \frac{\eta_k^\nu}{z}\sum_{j=n_c+1}^{\mu}\hat{K}_j^\phi, \tag{125}$$

$$\eta_k^\nu := \begin{cases} 1 & \text{if } n_c + 1 \le k \le \nu \\ 0 & \text{if } 1 \le k \le n_c \end{cases}. \tag{126}$$

Unfortunately, $\hat{F}_k^\phi$ and $\hat{F}_k^\theta$ are not Hermitian. Thus, we cannot easily transform the EHF equations of Eqs. (121)–(124) to the canonical forms [16,24]. However, like the canonical ROHF and ROHF-DODS (different orbitals for different spins) cases [48–52], if we properly and uniquely convert $\hat{F}_k^\phi$ and $\hat{F}_k^\theta$ to Hermitian operators, we would obtain the canonical EHF equations.

## 9. Conclusions

Pons Viver [22] proposed that the complicated EHF formulas including the multiple summation such as Eqs. (8), (25), and (48) can be compactly expressed by using the property of characteristic polynomials (Eq. (16)). In the present study, we have proven that the formulas $\mathcal{F}(z)$ and $\mathcal{V}(z)$ proposed by Pons Viver are mathematically equivalent to the original EHF formulas including the multiple summation when the condition of Eq. (37) is satisfied. Interestingly, for open-shell molecules the condition requires the use of ROHF-like orbitals [43] in calculating $\mathcal{F}(z)$ and $\mathcal{V}(z)$. That is, due to the condition, we cannot calculate $\mathcal{F}(z)$ of Eq. (28) and $\mathcal{V}(z)$ of Eq. (51) by using both UHF orbitals and extended pairing theorem, while we can calculate $\mathcal{G}(z)$ of Eq. (20) by using UHF orbitals.

Moreover, as shown by Pons Viver [22], by using the CO model, we have obtained the surprisingly compact expression of the EHF energy (please compare Eq. (99) and Eq. (111)). However, we have found that the EHF equations based on the CO model (i.e., the CO-based EHF equations) do not take the canonical forms, like the ROHF method, if we do not perform any additional transformation. Incidentally, even if the CO model is not employed, it is known that the EHF equations do not take the canonical forms [10–12]. As a result, Mayer et al. transformed the EHF equations to some useful forms [10–12]. In future work we would like to elucidate the relation between the canonical ROHF [48–52] and the CO-based canonical EHF methods, although we must first derive the CO-based canonical EHF equations.